\RequirePackage{ifpdf}
\documentclass[a4paper,11pt,hyper]{JHEP3}
\usepackage{amsmath,amsfonts,amssymb,color,graphicx,latexsym,theorem,mathrsfs,slashed,cite}
\usepackage{color}

\newcommand{\ma}[1]{\mbox{$\mathcal{#1}$}}
\newcommand{\mas}[1]{\mbox{$\mathscr{#1}$}}

\newcommand{\D}{{\rm d}}
\newcommand{\ti}{\tilde}
\newcommand{\we}{\wedge}

\title{
Geometry of Killing spinors in neutral signature
}

\author{Dietmar Klemm and Masato Nozawa\\
Universit\`a di Milano and INFN, Sezione di Milano,
Via G. Celoria 16, 20133 Milano, Italia}

\abstract{ 
We classify the supersymmetric solutions of minimal $N=2$ gauged supergravity in four
dimensions with neutral signature. They are distinguished according to the sign of the cosmological constant and whether the vector field constructed as a bilinear of the Killing spinor is null or non-null. In neutral signature the bilinear vector field can be spacelike, which is a new feature not arising in Lorentzian signature. In the $\Lambda<0$ non-null case, the canonical form of the metric is described by a fibration over a three-dimensional base space that has $\text{U}(1)$ holonomy with torsion.
We find that a generalized monopole equation determines the twist of the bilinear Killing field, which is
reminiscent of an Einstein-Weyl structure. If, moreover, the electromagnetic field strength is self-dual,
one gets the Kleinian signature analogue of the Przanowski-Tod class of metrics, namely a
pseudo-hermitian spacetime determined by solutions of the continuous Toda equation, conformal
to a scalar-flat pseudo-K\"ahler manifold, and admitting in addition a charged conformal Killing spinor.
In the $\Lambda<0$ null case, the supersymmetric solutions define an integrable null K\"ahler structure. In the $\Lambda>0$ non-null case, the manifold is a fibration over a Lorentzian Gauduchon-Tod base space.
Finally, in the $\Lambda>0$ null class, the metric is contained in the Kundt
family, and it turns out that the holonomy is reduced to ${\rm Sim}(1)\times{\rm Sim}(1)$. There appear
no self-dual solutions in the null class for either sign of the cosmological constant. 
}

\preprint{IFUM-1037-FT}

\begin{document}

\section{Introduction}

In the past decade, the program to systematically obtain geometries admitting Killing spinors has been intensively developed motivated by string theory. Supersymmetric solutions are characterized by the existence of at least one Killing spinor obeying a certain kind of first order differential equations.
The bilinear tensor fields built out of the Killing spinor define a privileged $G$-structure, which reduces the ${\rm Spin}(D-1,1)$ frame bundle to a subbundle $G$. The $G$-structures restrict tightly the geometry and the fluxes according to the torsion class \cite{Gauntlett:2001ur,Gauntlett:2002sc}. Generalizing earlier work by Tod~\cite{Tod:1983pm}, the seminal paper by Gauntlett et.~al \cite{Gauntlett:2002nw} has triggered
many new developments in this field, cf.~\cite{Gauntlett:2003fk,Gutowski:2004yv,Gutowski:2005id,
Bellorin:2006yr,Bellorin:2007yp,Gutowski:2003rg,Gauntlett:2002fz,Gauntlett:2003wb,Caldarelli:2003pb,
Bellorin:2005zc,Nozawa:2010rf,Meessen:2010fh,Meessen:2012sr} for an (incomplete) list of references.
Due to the reduction of the equations of motion to a simpler set of equations on a certain base space
of reduced holonomy, over which the full spacetime is fibered,
now a huge catalogue of supersymmetric solutions is available, some of which have been missed in the old
ansatz-based approach. In addition to the interest in these supersymmetric backgrounds in their own right,
they have many fruitful applications in holography and phenomenological model building based on flux
compactifications.

There exists another seminal work by Gillard, Gran and  Papadopoulos \cite{Gillard:2004xq}, who used the
so-called spinorial geometry technique. The basic idea behind this approach is to express spinors in terms
of differential forms and to use the gauge symmetry to transform them to a preferred representative of
their orbit. In this way the Killing spinor equations boil down to a linear system that can be used to
determine the metric and the other fields. This method turns out to be particularly adapted to geometries
admitting more than one Killing spinor. Moreover, it led to remarkable progress in constructing a large
variety of supersymmetric solutions~\cite{Gran:2005wn,Gran:2005ct,Cacciatori:2007vn,
Gutowski:2007ai,Grover:2008ih,Cacciatori:2008ek,Klemm:2009uw,Klemm:2010mc}. 

Apart from motivations coming from string theory, supersymmetric solutions have fundamental connections and impacts in mathematics. This includes the fields of special holonomy, generalized calibrations, integrable systems, complex manifolds and twistor spaces. In particular, the classification program of supersymmetric solutions resembles that of instantons and monopoles in gauge theories. This feature is more manifest in
non-Lorentzian manifolds. The signature of the metric affects the geometry in some crucial ways, the most
prominent example being perhaps the existence of solutions with self-dual Maxwell field and/or Weyl
tensor. In Lorentzian signature, we do not have counterparts of these solutions. In
\cite{Dunajski:2010zp,Gutowski:2010zs,Dunajski:2010uv}, Euclidean supersymmetric solutions have been classified and it turns out that these geometries enjoy much richer mathematical properties than
non-self-dual ones. Note that Euclidean supersymmetric solutions have additional applications in
localization techniques which allow to exactly compute the partition function of some
Euclidean superconformal field theories, that can then be compared with the result obtained from the
gravity side \cite{Martelli:2012sz}. 

In this paper, we shall be interested in supersymmetric solutions in neutral signature ($-,+,+,-$) by
focusing on the Wick-rotated version of minimal $N=2$ gauged supergravity. Thus far, not much has been
done on supersymmetric solutions and geometries in neutral (also called Kleinian or ultrahyperbolic)
signature (for some notable exceptions cf.~\cite{Bryant:2000,Dunajski:2001ea,Dunajski:2006un,
Dunajski:2006mk,Hervik:2012zz}), which is in some respect close to the Euclidean case, since field strengths can be
(anti-)self-dual and analogues of Hermitian and K\"ahler manifolds exist. Moreover, also a null class
of solutions appears, which is absent in Euclidean signature. There is thus a rich mathematical
structure to be explored. For instance we shall see that the null class of solutions admits an integrable
null K\"ahler structure, which is intrinsic to neutral signature. We will study these characteristic aspects
and clarify the underlying abundant mathematical structures.

Although considering Kleinian signature might seem a purely mathematical problem, there are also
several physical reasons that motivate this. First of all, two-time physics (cf.~\cite{Bars:2000qm} for a
review) has interesting applications in various areas, like cosmology \cite{Bars:2013yba} or
M-theory \cite{Bars:1999nk}. Moreover, Ooguri and Vafa \cite{Ooguri:1990ww} showed that the critical
dimension of the $N=2$ superstring is four, and then computed some scattering amplitudes,
which indicated that the bosonic part of the $N=2$ theory corresponds to self-dual metrics of
ultrahyperbolic signature ($-,+,+,-$). Let us finally mention that $(2,2)$-signature is intimately
related to twistor space \cite{Penrose:1967wn}, which is an important tool in perturbative computations
of scattering amplitudes in gauge theories \cite{Witten:2003nn}.

The present paper is organized as follows. In the next section, we fix our notations and describe the
minimal $N=2$ gauged supergravity theory on which we focus. In the following two sections, we tackle
the classification program of supersymmetric solutions depending on the sign of the cosmological constant;
the $\Lambda<0$ case in section~\ref{sec:AdS} and the $\Lambda>0$ case in section~\ref{sec:dS}. 
Section~\ref{sec:summary} summarizes the paper and points out some possible future work. 
Several appendices supplement the body of the text. 

\section{Minimal $N=2$ gauged supergravity}
\label{sec:N2SUGRA}

We shall consider a four-dimensional manifold endowed with a metric $g_{\mu\nu}$ of neutral signature,
i.e.,  $g_{\mu\nu}$ has two positive and two negative eigenvalues. 
The Einstein-Maxwell theory with a cosmological constant is described by the 
action 
\begin{align}
\label{action}
 S=\frac{1}{16\pi G} \int (R-2 \Lambda ) \star 1  -2 F \we \star F  \,, 
\end{align}
where $F$ is the Faraday tensor and $\Lambda$ is the cosmological constant.   
The bosonic equations of motion derived from the action read 
\begin{align}
\label{}
R_{\mu\nu}= \Lambda g_{\mu\nu}+2 \left(F_{\mu\rho}F_\nu{}^\rho -\frac 14
g_{\mu\nu}F_{\rho\sigma}F^{\rho\sigma} \right) \,, \qquad 
\D \star F=0 \,, \qquad \D F=0\,. 
\end{align}
The last equation can be solved locally in terms of the vector potential as 
$F=\D A$. 

A bosonic solution to this system is said to be supersymmetric 
if it admits a spinor $\epsilon$ satisfying 
\begin{align}
\hat\nabla_\mu \epsilon \equiv \left(\nabla_\mu +\frac{i}{4} F_{\nu\rho }\gamma^{\nu\rho }\gamma_\mu
-i\sqrt{-\frac{\Lambda}3}A_\mu +\frac{1}{2}\sqrt{-\frac{\Lambda}3} \gamma_\mu \right)\epsilon = 0\,. \label{KSE}
\end{align}
When $\Lambda$ is negative, the gauging is ${\rm U}(1)$, whereas the positive $\Lambda$
case corresponds to the noncompact $\mathbb R$-gauging. 

For Lorentzian signature, the sign of the Maxwell term in the action \eqref{action} is fixed by
requiring positivity of the kinetic energy. In neutral signature, there is a priori no reason to choose
the minus sign. For simplicity of our argument, we stick in this paper to the ordinary sign convention
above and do not attempt to consider a generalization to the plus sign. Instead, we will discuss in
appendix~\ref{sec:Int} how to construct the Killing spinor equation compatible with equations of motion in
such general settings. 

Let us summarize the convention of gamma matrices and fix our notation. 
The gamma matrices satisfy
\begin{align}
 \{\gamma_a , \gamma_b\}=2 \eta _{ab }=2{\rm diag}(-1,1,1,-1)_{ab} \,. 
\end{align}
$\gamma^0, \gamma^3$ are anti-hermitian, whence  we have 
\begin{align}
\gamma^\dagger_\mu=\gamma^0\gamma^3\gamma_\mu \gamma^3\gamma^0 \,. 
\end{align}
We define the chiral matrix by $\gamma_5 \equiv \gamma_{0123} $, 
yielding 
\begin{align}
 \gamma_5^\dagger =\gamma_5 \,, \qquad  
\gamma_{ab} =-\frac 12 \epsilon_{abcd}\gamma^{cd}\gamma_5 \,, \qquad 
\gamma_{abc}=\epsilon_{abcd}\gamma_5 \gamma^d\,. \label{gamma_eps}
\end{align}
Here $\epsilon_{abcd}$ is an alternate tensor with $\epsilon_{0123}=1$.

\subsection{Bilinear relations}

In this paper, we shall use the method of bilinears to classify all the supersymmetric solutions. Of course, the complementary spinorial geometry approach could also be applied. 

Suppose $\epsilon$ is a commuting ${\rm SO}(2,2)$ Dirac spinor. 
In terms of $\epsilon$, we can define the bilinear tensors \cite{Caldarelli:2003pb}
\begin{subequations}
\label{bilinears}
\begin{align}
E&\equiv\bar \epsilon \epsilon \,,\\
B&\equiv \bar \epsilon \gamma_5 \epsilon \,, \\ 
V_\mu &\equiv \bar \epsilon \gamma_\mu \epsilon \,, \\
U_\mu &\equiv  i\bar \epsilon \gamma_5 \gamma_\mu \epsilon \,,\\
\Phi_{\mu\nu }&\equiv i\bar \epsilon \gamma_{\mu\nu } \epsilon \,,
\end{align}
\end{subequations}
where the Dirac conjugation is defined by 
$\bar \epsilon \equiv -i\epsilon^\dagger \gamma^0\gamma^3$. 
This convention ensures that the above bilinears are all real. 
These tensorial quantities will play a central role in our analysis.
We first note that any $4\times 4 $ matrix $M$ can  be expanded in terms of 
a Clifford basis as 
\begin{align}
M=\frac 14 \left[{\rm Tr}(M) \mathbb I_4 
+{\rm Tr}(M\gamma_5)\gamma_5 +{\rm Tr}(M\gamma_\mu)\gamma^\mu  
-{\rm Tr}(M\gamma_5\gamma_\mu)\gamma_5\gamma^\mu  
-\frac 12 {\rm Tr}(M\gamma_{\mu\nu})\gamma^{\mu\nu}
\right]\,. \nonumber 
\end{align}
Viewing $\epsilon \bar \epsilon$ as a $4\times 4$ matrix, 
one can obtain various relations between bilinears using the above formula. 
A simple computation gives the projection relations
\begin{align}
iU^\mu \gamma_\mu  \epsilon =(-B+E\gamma_5 )\epsilon \,, \quad \,
V^\mu \gamma_\mu \epsilon =(E-\gamma_5 B)\epsilon \,, \quad \,
i\star \Phi_{\mu\nu}\gamma^{\mu\nu} \epsilon = -2 (B+\gamma_5 E) \epsilon \,, 
\label{fierz_projection}
\end{align}
which immediately imply
\begin{align}
 V^\mu V_\mu &=U^\mu U_\mu =E^2-B^2 \,, \qquad 
V^\mu U_\mu =0 \,, \qquad  \Phi_{\mu\nu }\Phi^{\mu\nu }= 2 (E^2+B^2)\,.
\end{align}
Moreover, it is straightforward to derive the additional relations
\begin{align}
 \Phi_{\mu\nu }V^\nu &=B U_\mu \,, \qquad \Phi_{\mu\nu }U^\nu =-B V_\mu \,, \label{alg_PhiUV}\\ 
\star \Phi_{\mu\nu }U^\nu &=E V_\mu \,, \qquad \star \Phi_{\mu\nu }V^\nu =-E U_\mu \,,
\label{alg_starPhiUV} \\
E \Phi_{\mu\nu }&=-B \star \Phi_{\mu\nu } +\epsilon_{\mu\nu\rho\sigma }V^\rho U^\sigma \,, 
\label{alg_EBPhi}\\
\Phi_{\mu\rho }\star \Phi_\nu {}^\rho &=\frac{1}{4} g_{\mu\nu }\Phi^{\rho\sigma }\star \Phi_{\rho\sigma }=-EB g_{\mu\nu} \,, \label{alg_PhistarPhi}\\
\Phi_{\mu\rho }\Phi_\nu {}^\rho &=-(V_\mu V_\nu +U_\mu U_\nu )+E^2 g_{\mu\nu } \,, 
\label{alg_Phisq}
\\
\star \Phi_{\mu\rho }\star \Phi_\nu {}^\rho &=V_\mu V_\nu +U_\mu U_\nu +B^2 g_{\mu\nu } \,,
\label{alg_starPhisq}
\end{align}
where 
$\star\Phi_{\mu\nu}=(1/2)\epsilon_{\mu\nu\rho\sigma}\Phi^{\rho\sigma}=-i\bar \epsilon\gamma_5\gamma_{\mu\nu}\epsilon$. 

In addition to the bilinears introduced in (\ref{bilinears}), 
it turns out to be convenient to define subsidiary tensorial quantities as 
\begin{align}
\label{WPsi}
W_\mu \equiv \epsilon ^T C^{-1}\gamma_\mu \epsilon \,, \qquad 
\Psi_{\mu\nu}\equiv {i}\epsilon ^T C^{-1}\gamma_{\mu\nu} \epsilon \,. 
\end{align}
Here $C$ is the charge conjugation matrix satisfying 
\begin{align}
C^{-1}\gamma_\mu C=-\gamma_\mu ^T \,, \qquad 
C^T=-C \,.
\end{align}
This gives the useful relation
$\star \Psi_{\mu\nu}=-i\epsilon^T C^{-1}\gamma_5\gamma_{\mu\nu}\epsilon$. 
Note that $W_\mu$ and $\Psi_{\mu\nu}$ are complex tensors. 
Expanding the matrix $\epsilon\epsilon^T C^{-1}$ in terms of a Clifford basis, 
we can get quartic relations between the bilinears in the same way as above. 
For instance one gets the orthogonality property  
\begin{align}
V^\mu W_\mu &=U^\mu W_\mu =W^\mu W_\mu =0 \,, \qquad 
W^\mu \bar W_\mu =2 (B^2-E^2) \,,
\end{align}
as well as
\begin{align}
(B^2-E^2)g_{\mu\nu} &=-V_\mu V_\nu-U_\mu U_\nu +W_{(\mu }\bar W_{\nu)} \,, 
\label{gbasis}\\
(B^2-E^2)\Psi_{\mu\nu}&= 2({i}EV_{[\mu}{-}B U_{[\mu})W_{\nu]} \,.
\label{Psirel}
\end{align}
Eq.~(\ref{gbasis}) implies that 
$(V_\mu, U_\mu , W_\mu, \bar W_\mu )$ form a complete basis 
when $B^2-E^2$ is nonvanishing, which is a main advantage of introducing the supplementary 
tensors (\ref{WPsi}). 

We would like to stress  that the sign of $B^2-E^2$ is left undetermined. This property is
in contrast to the case of Lorentzian or Euclidean signature, where 
the causal nature of the bilinear (pseudo)vectors is fixed. 

In the following sections, we divide our discussion according to the sign 
of $\Lambda$ and the causal nature of the vectors $V^\mu,U^\mu$.
We classify all the supersymmetric solutions and discuss their geometric properties.


\section{Negative $\Lambda$}
\label{sec:AdS}

Let us begin with the $\Lambda=-3\ell^{-2}<0$ case. Here 
$\ell$ is the `pseudo-AdS' curvature radius and the Killing spinor equation (\ref{KSE}) 
reduces to 
\begin{subequations}
\begin{align}
 \hat \nabla_\mu \epsilon &\equiv\left(\nabla_\mu +\frac{i}{4} F_{\nu\rho }\gamma^{\nu\rho }\gamma_\mu -\frac{i}{\ell}A_\mu 
+\frac{1}{2 \ell}\gamma_\mu \right)\epsilon = 0\,, \\
\overline{\hat \nabla_\mu \epsilon}&=\overline{\nabla_\mu\epsilon}
+\bar \epsilon \left[\frac{i}2 (F_{\mu\nu}+\star F_{\mu\nu}\gamma_5)\gamma^\nu 
+\frac{i}\ell A_\mu +\frac 1{2\ell}\gamma_\mu \right] \,, \\
\hat \nabla_\mu (\epsilon^TC^{-1}) &=\nabla_\mu (\epsilon^TC^{-1})
+\epsilon^TC^{-1} \left[\frac i2 (F_{\mu\nu}+\star F_{\mu \nu}\gamma_5 )\gamma^\nu 
-\frac i\ell A_\mu -\frac 1{2\ell}\gamma_\mu \right]\,.
\end{align}
\end{subequations}
With these at hand, one can derive the following linear differential relations for the bilinears:
\begin{subequations}
\begin{align}
\nabla_\mu E&=-\star F_{\mu\nu}U^\nu -\frac 1\ell V_\mu \,, \label{AdS_diff_E}\\
\nabla_\mu B&=F_{\mu\nu}U^\nu \,, \label{AdS_diff_B}\\
\nabla_\mu V_\nu &= -\frac 1\ell E g_{\mu\nu}+F_{(\mu}{}^\rho\Phi_{\nu)\rho}
-\star F_{(\mu}{}^\rho \star \Phi _{\nu)\rho}\,, \label{nabla_V} \\
\nabla_\mu U_\nu &= -\frac 1\ell \star \Phi_{\mu\nu}-B F_{\mu\nu}-E \star F_{\mu\nu}\,, \label{nabla_U} \\
\nabla_\mu \Phi_{\nu\rho}&= -\frac 1\ell \epsilon_{\mu\nu\rho\sigma}U^\sigma +2 F_{\mu [\nu}
V_{\rho]}-V_\mu F_{\nu\rho}-2 g_{\mu[\nu}F_{\rho]\sigma} V^\sigma \,,
\label{AdS_Phieq}
\end{align}
\end{subequations}
and 
\begin{subequations}
\begin{align}
\nabla_\mu W_\nu & =-\frac {i}\ell \Psi_{\mu\nu}+\frac {2i}\ell A_\mu W_\nu 
{+}F_{(\mu}{}^\rho \Psi_{\nu)\rho} -\star F_{(\mu }{}^\rho \star \Psi_{\nu)\rho} \,, \label{nabla_W} \\ 
\nabla_\mu \Psi_{\nu\rho }&= \frac{ 2i}\ell g_{\mu [\nu}W_{\rho]}{-}\frac {2}\ell A_\mu \Psi_{\nu\rho}
+2 F_{\mu[\nu}W_{\rho ]} -W_\mu F_{\nu \rho}-2 g_{\mu[\nu}F_{\rho ]\sigma}W^\sigma\,. 
\end{align}
\end{subequations}
It follows from (\ref{nabla_U}) that $U^\mu$ is a Killing vector,
\begin{align}
\mas L_U g_{\mu\nu}=0 \,. 
\end{align}
Provided the Maxwell equation $\D \star F=0$ and the Bianchi identity $\D F=0$ hold, 
the differential relations (\ref{AdS_diff_E}) and (\ref{AdS_diff_B}) imply that the Maxwell 
field is also invariant under the action of $U$, 
\begin{align}
\label{}
\mas L_U F= 0 \,, \qquad \mas L_U \star F=0 \,. 
\end{align}
In the following we will obtain the local form of the metric depending on whether the Killing field $U^\mu$
is null or not. We refer to the former as the null class, and the latter as the non-null class.

\subsection{Non-null class}

Assuming that $f\equiv B^2-E^2 $ is nonvanishing, eqs.~(\ref{alg_PhiUV}), (\ref{alg_starPhiUV}) imply
that $\Phi$ can be solved in terms of the other bilinears as 
\begin{align}
\Phi_{\mu\nu }=\frac 1 f \left(
2 B V_{[\mu}U_{\nu]} -E \epsilon_{\mu\nu\rho\sigma}V^\rho U^\sigma \right)\,. \label{Phi-EBUV}
\end{align}
Similarly, the differential relations for $E$ and $B$ give the Maxwell field
\begin{align}
F_{\mu\nu}=\frac 1f \left[2 U_{[\mu }\nabla_{\nu]}B-\epsilon_{\mu\nu\rho\sigma}
U^\rho (\nabla^\sigma E +\ell^{-1} V^\sigma)\right] \,. 
\label{Feq}
\end{align}
It follows then that the equation (\ref{AdS_Phieq}) for $\Phi_{\mu\nu}$ automatically follows from the other
differential constraints. 

Since $U^\mu $ is a Killing field, it is convenient to introduce a coordinate system 
in such a way that $ U^\mu$ is a coordinate vector, $U=\partial/\partial t$, and the 
metric takes a $t$-independent form, 
\begin{align}
\label{metric_AdS}
\D s^2=-f(\D t +\omega )^2 +f^{-1} h_{mn}\D x^m \D x^n \,.
\end{align}
Here the one-form $\omega$ measures the twist of the vector field $U$ and $f^{-1} h_{mn}$
is the Lorentzian base space metric orthogonal to $U$. We have added the prefactor $f^{-1}$ 
for convenience so that $h_{mn}$ describes the three-dimensional Einstein frame metric when 
one performs a Kaluza-Klein reduction along $t$.

Let us next introduce a local coordinate system on the base space. To this end, 
we first notice that the relation (\ref{Psirel}) implies that $\Psi_{\mu\nu}$ is also redundant 
when $f$ is nonvanishing. 
Inserting \eqref{Psirel} into \eqref{nabla_W} and using (\ref{Feq}), we get 
\begin{align}
\label{dWeq}
\D W=-\frac{2i}\ell \left(\frac{i E V-B U}f -A \right)\we W \,. 
\end{align}
The differential relations for $V$ (\ref{nabla_V}) and $W$ (\ref{dWeq}) therefore imply 
\begin{align}
\D V= 0 \,, \qquad W \we \D W =0 \,,
\end{align}
hence $V$ is closed and $W$ is hypersurface-orthogonal.  
Choosing the phase of the Killing spinor appropriately, 
one can thus introduce local scalars ($x, y, z$) and $\phi$ by
\begin{align} 
\label{local_scal}
V_\mu =\nabla_\mu z \,, \qquad W_\mu =e^\phi (\nabla_\mu x +i\nabla_\mu y ) \,, 
\end{align}
with 
\begin{align}
h_{mn}\D x^m \D x^n =- \D z^2 + e^{2\phi}(\D x^2+\D y^2) \,. \label{metric_base}
\end{align}
Here $\phi=\phi(x,y,z)$ is a function on the base space.  
In appendix~\ref{app_holonomy}, we determine the holonomy of this base space. 

Let us next look at the symmetric part of (\ref{nabla_V}). 
Introducing  Maxwell potentials by 
\begin{align}
F_\pm \equiv  \frac{2}{\ell (E\pm B )}\,, 
\end{align}
the only constraint arising from (\ref{nabla_V}) is a first-order differential equation for $\phi$,
\begin{align}
\phi '=-\frac 12 (F_++F_- ) \,, 
\label{eq_phip}
\end{align}
where the prime denotes a partial derivative with respect to $z$. This is a
restriction describing the embedding of the two-surface $e^{2\phi}(\D x^2+\D y^2)$
into the base space $\D s^2(h)=h_{mn}\D x^m\D x^n$. 

Imposing the Maxwell equations and Bianchi identity on (\ref{Feq}), 
we get the two decoupled equations 
\begin{align}
\Delta F_\pm - e^{2\phi }(F_\pm ^3-3 F_\pm F_\pm ' +F_\pm '')=0 \,, 
\label{MaxBia_Fpm}
\end{align}
where $\Delta=\partial_x^2+\partial_y^2$ denotes the two-dimensional flat Laplacian. 
Viewing $U=-f (\D t+\omega)$ as a 1-form, the differential relation for $U$ 
yields the governing equation for the base space 1-form $\omega$,
\begin{align}
\nabla_{[\mu } \omega _{\nu]} =-\frac 1{2f^2}\epsilon_{\mu\nu\rho\sigma}
U^\rho \Omega^\sigma\,, \label{ext-der-omega}
\end{align}
where $\Omega_\mu$ measures the twist of the Killing vector,
\begin{align}
\label{AdS_Omega_def}
\Omega_\mu \equiv \epsilon_{\mu\nu\rho\sigma} U^\nu \nabla^\rho U^\sigma
= 2 (B \nabla_\mu E -E \nabla_\mu B+2\ell^{-1} B V_\mu ) \,. 
\end{align}
Here we have used (\ref{nabla_U}) in the second step.
Written down explicitly, \eqref{ext-der-omega} reads
\begin{align}
\label{eq_domega}
\partial_x \omega_y -\partial_y \omega_x&=f^{-2} e^{2\phi}\Omega_z \,, \nonumber \\
\partial_y \omega_z-\partial_z \omega_y&=-f^{-2} \Omega_x \,, \\
\partial_z \omega_x-\partial_x \omega_z&=-f^{-2} \Omega_y\,.\nonumber 
\end{align}
The integrability condition of this equation is assured by the Maxwell equations
and Bianchi identity (\ref{MaxBia_Fpm}). We shall come back to this point more in detail below.

Let us finally obtain the equation that determines $\phi$. To this end, 
we employ the gauge
\begin{align}
U^\mu A_\mu =B \,, \label{gauge}
\end{align}
which implies that the gauge potential $A_\mu$ is also time-independent, $\mas L_U A_\mu=0 $. 
Plugging (\ref{local_scal}) into (\ref{dWeq}) and using (\ref{eq_phip}), 
we find 
\begin{align}
\label{calBeq}
\ma B_z=0 \,, \qquad \partial_x \phi=-\frac 2\ell \ma B_y \,, \qquad 
\partial_y \phi =\frac 2\ell \ma B_x \,, 
\end{align}
where $\ma B_m\equiv A_m -B \omega_m$. 
This allows us to obtain the gauge potential 
\begin{align}
\label{}
A=B(\D t+\omega )+\frac \ell 2(\partial_y \phi \D x-\partial_x \phi \D y)\,. 
\label{Aeq}
\end{align}
The compatibility condition $F=\D A$ of (\ref{Feq}) and (\ref{Aeq}) 
yields 
\begin{align}
\label{delta_phi}
\Delta \phi +\frac 12 e^{2\phi}[F_+'+F_-'-F_+^2-F_-^2+F_+F_-]=0 \,. 
\end{align}
This is equivalent to the trace part of Einstein's equations, 
provided that eqs.~(\ref{eq_phip}), 
(\ref{MaxBia_Fpm}) and  (\ref{eq_domega}) are satisfied. 

We have exhausted the bilinear equations. The above bosonic configurations 
are obviously necessary constraints for the preservation of supersymmetries. 
We shall now show that these are also sufficient.  
Let us take the tetrad frame, 
\begin{align}
e^0 = f^{1/2}(\D t +\omega )\,, \qquad e^i =f^{-1/2} \hat e^i \,, 
\end{align}
where $\hat e^i$ is an orthonormal frame for the base space, 
\begin{align}
\hat e^1= e^\phi  \D x \,, \qquad \hat e^2=e^\phi \D y \,, \qquad 
\hat e^3 =\D z\,. 
\end{align}
Using the formula for the Lie derivative of a spinor field along a (conformal) Killing vector, 
the time-independent spinor $\mas L_U\epsilon=\partial_t \epsilon=0$  solves the time component of the Killing spinor equation  $U^\mu \hat \nabla_\mu \epsilon=0$ 
under the gauge condition (\ref{gauge}) as
\begin{align}
\mas L_U \epsilon &\equiv U^\mu \nabla_\mu \epsilon +\frac 14 \nabla_\mu U_\nu \gamma^{\mu\nu} 
\epsilon \nonumber \\
&=U^\mu \hat\nabla_\mu \epsilon+\frac{i}{\ell} (U^\mu A_\mu -B)\epsilon  =0\,, 
\end{align}
where we have used the projection condition (\ref{fierz_projection}) for 
$U^\mu$ and $\star \Phi_{\mu\nu}$. Using the expressions given in appendix~\ref{app:spincon}, 
the base space component of the Killing spinor equation reads 
\begin{align}
\left[D_m +\frac 1{2f}(\partial_m B+\gamma_5 E)(-B+\gamma_5 E)-\frac i\ell \ma B_m 
+\frac{E}{\ell f^{3/2}}\hat \gamma_{mn} V^n \right]\epsilon =0\,, \label{KSE-base}
\end{align}
where $\ma B_m \equiv A_m -B \omega_m$ as before and $D_m$ denotes the Levi-Civita
connection of the base space metric $h_{mn}$. 
Decomposing $\epsilon=\sqrt{B+E}\zeta^++\sqrt{B-E}\zeta^-$
with $\gamma_5 \zeta^\pm =\pm \zeta^\pm$, \eqref{KSE-base} decouples into
\begin{align}
\left(D_m-\frac i\ell \ma B_m +\frac{E}{\ell f^{3/2}}\hat \gamma_{mn}V^n\right) \zeta^\pm =0 \,. 
\end{align}
Using the projection condition $i \gamma_{12}\epsilon =-\epsilon$ as well as (\ref{calBeq}), 
we find $\partial_m \zeta^\pm=0$, i.e, $\zeta^\pm$ are constant spinors. 
Taking into account $\gamma^0\zeta^+=-\zeta^- $, the solution therefore reads 
\begin{align}
\epsilon =
\left(\sqrt{B+E}-i \gamma^0 \sqrt{B-E} \right)(1-i \gamma_{12})(1+\gamma_5)\epsilon_0 \,, 
\end{align}
where $\epsilon_0$ is a constant Dirac spinor. 
It follows that the spacetime preserves at least the fraction of 1/4 supersymmetry. 

This illustrates the use of the subsidiary bilinear fields ($W_\mu, \Psi_{\mu\nu}$). 
Ref.~\cite{Caldarelli:2003pb} worked only with the Maxwell field strength and the 
gauge potential was not obtained explicitly. In that case, (\ref{delta_phi}) was derived by requiring the integrability condition of the Killing spinor and the final solution for the Killing spinor has an additional
phase. The simplification achieved in this paper is not obtainable without introducing (\ref{WPsi}). 

To summarize, the geometry of Killing spinors with $\Lambda<0$ can be specified by 
solving the nonlinear system (\ref{eq_phip}), (\ref{MaxBia_Fpm}) and (\ref{delta_phi}). 
This coupled system is very similar to its Lorentzian counterpart given in \cite{Caldarelli:2003pb}. 
A notable feature of the neutral signature is that the norm of the Killing vector $U^\mu$ can take
either sign. This does not matter at all since the neutrality of the metric is preserved under the reflection
$f\mapsto -f$. This property allows for a richer class of supersymmetric solutions than in the
Lorentzian or Euclidean cases. 
Note that $g_{\mu\nu}\mapsto-g_{\mu\nu}$ under $f\mapsto -f$. In string theory, this symmetry is
associated with what is usually referred to as `crossing symmetry'. In \cite{Barrett:1993yn}, it was termed
`chronal-chiral symmetry'.

One can easily check that similar to the Lorentzian case \cite{Cacciatori:2004rt}, 
eqs.~(\ref{eq_phip}), (\ref{MaxBia_Fpm}) and (\ref{delta_phi}) are invariant 
under the ${\rm PSL}(2,\mathbb R)$ transformations 
\begin{align}
\label{SL2R}
z\mapsto \frac{az+b}{cz+d}\,, \qquad ad-bc=1 \,, 
\end{align}
provided that $F_\pm $ and $\phi$ transform as
\begin{align}
\label{SL2R-Fphi}
F_\pm \mapsto (cz+d)^2 F_\pm +\partial_z[(cz+d)^2] \,, \qquad \phi \mapsto 
\phi-2 \log (cz+d) \,, 
\end{align}
Under \eqref{SL2R}, \eqref{SL2R-Fphi}, the base space is conformally rescaled as
$h_{mn} \mapsto (cz+d)^{-4} h_{mn}$. This transformation preserves supersymmetry, 
but maps nontrivially the solution into a new one. One might ask whether this ${\rm PSL}(2,\mathbb R)$
is related to the Ehlers transformations for the (electro)vacuum solutions to Einstein's equations.
The latter symmetry is, however, broken in the presence of a cosmological constant \cite{Klemm:2015uba}.
Thus, the remarkable symmetry (\ref{SL2R}) is not related to Ehlers transformations and is intrinsic to supersymmetric solutions.

The Bianchi identity part of (\ref{MaxBia_Fpm}) is yet actually redundant, since it automatically follows from the other equations (this is obvious since (\ref{delta_phi}) arises from the compatibility condition $F=\D A$).
To further reduce the governing equations, we define 
\begin{align}
\label{}
\mas B\equiv \frac 12 (F_+-F_-) \,. 
\end{align}
Then, \eqref{eq_phip}, \eqref{MaxBia_Fpm} and \eqref{delta_phi} boil down to
\begin{subequations}
\label{BPhieq}
\begin{align}
\Delta \phi-\frac 12 e^{2\phi}(2\phi''+\phi'{}^2+3 \mas B^2)&=0 \,, 
\label{BPhieq1}\\
\Delta \mas B -e^{2\phi} (\mas B^3+\mas B'' +3\mas B'\phi'+3\mas B\phi'{}^2+3 \mas B \phi'')&=0\,.
\end{align}
\end{subequations}
These equations can be derived from the three-dimensional action
\begin{align}
\label{}
S_3 = \int \D ^2 x \D z \left[
\nabla \mas B \cdot \nabla \phi+\frac 12 e^{2\phi}
(\mas B^3-2\mas B' \phi'-3 \mas B \phi'{}^2 )
\right] \,. 
\end{align}
(\ref{BPhieq}) also imply the conservation law
\begin{align}
\label{cons_law}
0=\rho '+\partial_i j_i  \,, 
\end{align}
where 
\begin{equation}
\rho = e^{2\phi}[(-\phi'{}^2+\mas B^2)\mas B-\phi' \mas B' +\mas B \phi '']\,, \qquad
j_i = \phi' \partial_i\mas B -\mas B \partial_i \phi ' \,. 
\end{equation}
Note that the conservation law (\ref{cons_law}) is a direct consequence of the 
4-dimensional identity $\nabla_\mu (f^{-2} \Omega^{\mu})\equiv 0$ 
for the twist (\ref{AdS_Omega_def}) of a Killing vector.

\subsubsection{Generalized monopole equation}

Let us return to the equation (\ref{eq_domega}) determining $\omega $. 
This can actually be written as a generalized monopole equation \cite{Jones:1985},
\begin{align}
\D\omega = \star_h \left(\D\Sigma + \frac12\nu\Sigma\right)\,, \label{gen_monopole}
\end{align}
where the function $\Sigma$ and the one-form $\nu$ are respectively given by
\begin{align}
\Sigma = -\frac{2EB}{(E^2-B^2)^2}\ln\left|\frac{E}{B}\right|\,, \qquad \nu_m =
\frac{4V_m}{E\ell\ln\left|\frac{E}{B}\right|} - 2\partial_m\ln\left|\frac{2EB}{(E^2-B^2)^2}\right|\,,
\end{align}
while $\star_h$ denotes the Hodge star with respect to the base space metric (\ref{metric_base}).
Notice also that the generalized monopole equation (\ref{gen_monopole}) is invariant under
Weyl rescaling, accompanied by a gauge transformation of $\nu$,
\begin{align}
h_{mn}\D x^m\D x^n \mapsto e^{2\psi}h_{mn}\D x^m\D x^n\,, \qquad \Sigma \mapsto
e^{-\psi}\Sigma\,, \qquad \nu \mapsto \nu + 2\D\psi\,. \label{Weyl-transf}
\end{align}
It would be very interesting to better understand the deeper origin of the conformal invariance
of (\ref{eq_domega}), which remains rather obscure in this context, since (unlike the self-dual subcase
and the $\Lambda>0$ class that will both be considered below) it is unclear if the base manifold
(\ref{metric_base}) is Einstein-Weyl. We will have more to say on this point in appendix \ref{app_holonomy}.

Note that the symmetry (\ref{Weyl-transf}) is not enhanced to an invariance of the full set of equations.
To see this, note that (\ref{Weyl-transf}) arises from the transformations 
$(E, B, V_m, \phi) \mapsto (e^{\psi/2}E, e^{\psi/2} B, e^{\psi}V_m, \phi+\psi)$.
From the four-dimensional point of view, this amounts to the conformal rescaling
$g_{\mu\nu}\mapsto e^\psi g_{\mu\nu}$. Obviously, this new form of the metric does not fall into the
canonical form (\ref{metric_AdS}). This means that the transformation (\ref{Weyl-transf}) does not preserve
supersymmetry. 

The integrability conditions for (\ref{gen_monopole}),
\begin{align}
\D\star_h\left(\D\Sigma + \frac12\nu\Sigma\right) = 0\,,
\end{align}
can be rewritten as
\begin{align}
\frac1{\sqrt{-h}}\partial_n\left[\sqrt{-h}h^{np}\left(\partial_p + \frac12\nu_p\right)\Sigma\right] = 0\,,
\label{int_cond}
\end{align}
or even more compactly as $\tilde D^2\Sigma=0$, with the Weyl-covariant derivative
\begin{align}
\tilde D_n \equiv D_n - \frac m2\nu_n\,,
\end{align}
where $D_m$ is the Levi-Civita connection of $h$ and $m$ denotes the Weyl weight of the
corresponding field\footnote{A field $\Gamma$ with Weyl weight
$m$ transforms as $\Gamma\mapsto e^{m\psi}\Gamma$ under a Weyl rescaling.}. It is straightforward
to show that in our case, (\ref{int_cond}) is equivalent to
\begin{align}
F_+\left[\Delta F_- - e^{2\phi}(F_-^3 - 3F_-F_-' + F_-'')\right] -
F_-\left[\Delta F_+ - e^{2\phi}(F_+^3 - 3F_+F_+' + F_+'')\right] = 0\,, \label{int_cond_eval}
\end{align}
if one uses in addition (\ref{eq_phip}). This is a linear combination of the two
eqs.~(\ref{MaxBia_Fpm}). Note that differentiation of (\ref{delta_phi}) w.r.t.~$z$ and subsequent use of
(\ref{eq_phip}) yields
\begin{align}
\Delta F_- - e^{2\phi}(F_-^3 - 3F_-F_-' + F_-'') +
\Delta F_+ - e^{2\phi}(F_+^3 - 3F_+F_+' + F_+'') = 0\,. \label{consequ_delta_phi}
\end{align}
Together, (\ref{int_cond_eval}) and (\ref{consequ_delta_phi}) imply (\ref{MaxBia_Fpm}). The actual
geometrical data are thus the generalized monopole equation (\ref{gen_monopole}) together with
(\ref{eq_phip}) and (\ref{delta_phi}). It would be interesting to see what the geometrical interpretation
of (\ref{delta_phi}) is. As it stands, it seems to be a sort of restriction on the (scalar) curvature of the base
space, but we were not able to figure out its precise meaning.

\subsubsection{Self-dual solution}
\label{selfdual-negLambda}

In this section, let us focus on the solution  with a self-dual field strength,
\begin{align}
\star F= F \,. 
\end{align}
In this case, the stress-energy tensor of the Maxwell field vanishes. 
Eqs.~(\ref{AdS_diff_E}) and (\ref{AdS_diff_B}) imply then
\begin{align}
B=-E-\frac{z}\ell \,,
\end{align}
where the integration constant has been set to zero by using the freedom $z\mapsto z+{\rm const}$. 
Introducing a new coordinate $w$,
\begin{align}
w=-\frac{\ell^2}z \,, 
\end{align}
and defining the new variables 
\begin{align}
\label{}
H=\left(1-\frac{2 E w}\ell \right)^{-1}  \,, \qquad 
e^u=\frac{w^4}{\ell^4}e^{2\phi} \,, 
\end{align}
the metric can be cast into the form
\begin{align}
\D s^2 =\frac{\ell^2}{w^2} \left[
-H^{-1} (\D t +\omega)^2 +H \{-\D w^2+e^u (\D x^2+\D y^2)\}
\right]\,. \label{metr_Prz_Tod}
\end{align}
\eqref{delta_phi} boils down to the hyperbolic continuous Toda equation 
\begin{align}
\Delta u -\partial_{w}^2 (e^u)=0 \,, \label{Todaeq}
\end{align}
while $H$ is given by
\begin{align}
H=\frac{w}2 \partial_{w} u -1\,.
\end{align}
One can verify that (\ref{metr_Prz_Tod}) is pseudo-Hermitian, with the pseudo-complex structure
$J$ given by
\begin{align}
{J_\mu}^\nu =  \frac{2\ell}{w}g^{\nu\lambda}\Phi^-_{\mu\lambda}\,,
\end{align}
where $\Phi^-\equiv\frac12(\Phi-\star\Phi)$ denotes the anti-self-dual part of the two-form $\Phi$.
Note that (\ref{nabla_U}) implies
\begin{align}
(\D U)^{\pm} = \mp\frac2{\ell}\Phi^{\pm} - 2(B\pm E)F^{\pm}\,.
\end{align}
Since $F^-$ vanishes in our case, $\Phi^-$ is proportional to the anti-self-dual part of the exterior
derivative of the Killing vector $U$. We have checked explicitly that the Nijenhuis tensor
of $J$ is zero, and thus the almost pseudo-complex structure $J$ is integrable. The results of this
subsection are actually the neutral signature analogue of the Euclidean case considered by
Przanowski \cite{Przanowski:1991ru} and Tod \cite{Tod:2006wj}\footnote{(\ref{metr_Prz_Tod}) falls
into class B of \cite{Przanowski:1991ru}.}. Moreover, if we define $\D\hat s^2$ by
$\D\hat s^2\equiv (w^2/\ell^2)\D s^2$, the resulting metric $\D\hat s^2$ is scalar-flat
pseudo-K\"ahler, with pseudo-K\"ahler form $\hat J$ given by
\begin{align}
\hat J_{\mu\nu} = \frac{w^2}{\ell^2} J_{\mu\nu}\,.
\end{align}
It is again a straightforward matter to check that $\hat\nabla_{\mu}\hat J_{\nu\rho}=0$, with
$\hat\nabla$ the Levi-Civita connection of $\D\hat s^2$. The latter is the Kleinian signature
version of LeBrun's class of scalar-flat K\"ahler metrics \cite{LeBrun:1991}.

Notice that the (anti-)self-dual part of $\Phi$ can be expressed as a bilinear of chiral spinors,
\begin{align}
\Phi^{\pm}_{\mu\nu} = i\bar\epsilon_{\mp}\gamma_{\mu\nu}\epsilon_{\mp}\,,
\end{align}
where $\epsilon_{\mp}=\frac12(1\mp\gamma_5)\epsilon$ satisfies
$\gamma_5\epsilon_{\mp}=\mp\epsilon_{\mp}$. In the case $F^{\pm}_{\mu\nu}=0$,
$\epsilon_{\mp}$ turns out to be a charged conformal Killing spinor (CCKS). To see this, start from
the Killing spinor equation (\ref{KSE}) and multiply from the left with the projector
$\Pi_{\mp}=\frac12(1\mp\gamma_5)$, which leads to
\begin{align}
\nabla_\mu\epsilon_{\mp} + \frac i4\slashed{F}\gamma_\mu\epsilon_{\pm} - \frac i{\ell}A_\mu
\epsilon_{\mp} + \frac1{2\ell}\gamma_\mu\epsilon_{\pm} = 0\,. \label{projected_KSE}
\end{align}
Now, using the second relation of (\ref{gamma_eps}), one shows that
\begin{align}
\slashed{F}\gamma_\mu\epsilon_{\pm} = F^{\pm}_{\rho\sigma}\gamma^{\rho\sigma}\gamma_\mu
\epsilon\,,
\end{align}
so that $\slashed{F}\gamma_\mu\epsilon_+=0$ for $F^+_{\rho\sigma}=0$ and analogous for the
minus sign. In the (anti-)self-dual case, (\ref{projected_KSE}) becomes therefore
\begin{align}
\nabla_\mu\epsilon_{\mp}  - \frac i{\ell}A_\mu\epsilon_{\mp} +
\frac1{2\ell}\gamma_\mu\epsilon_{\pm} = 0\,. \label{projected_KSE_asd}
\end{align}
Contracting this from the left with $\gamma^\mu$ gives
\begin{align}
\epsilon_{\pm} = -\frac{\ell}2\left(\slashed{\nabla} - \frac i{\ell}\slashed{A}\right)\epsilon_{\mp}\,,
\end{align}
which can be plugged back into (\ref{projected_KSE_asd}) to obtain
\begin{align}
\left[\nabla_\mu - \frac14\gamma_\mu\slashed{\nabla} - \frac i{\ell}\left(A_\mu - \frac14\gamma_\mu
\slashed{A}\right)\right]\epsilon_{\mp} = 0\,, \label{CCKS}
\end{align}
which is the charged conformal Killing spinor equation. (\ref{CCKS}) has been considered by
mathematicians before, see e.g.~\cite{Lichnerowicz:1989}, part III. In particular, it was shown in
\cite{Lischewski:2014ffa} (theorem 18) that in (2,2) signature a CCKS half spinor of nonzero length
equivalently characterizes the existence of pseudo-K\"ahler metrics in the conformal class.

\subsubsection{Reissner-Nordstr\"om-Taub-NUT family}

Let us give an example of a simple class of supersymmetric solutions. To this end, 
we decompose $\phi=\phi(x,y,z)$ into two contributions,
\begin{align}
\label{}
\phi (x,y,z) = \Phi_0(x,y,z)+\Xi(x,y) \,, \qquad \Phi_0(x,y,z) \equiv \int \D z  \partial_z\phi (x,y,z)\,, 
\end{align}
Assume that $\Phi_0$, $\mas B$ depend only on the coordinate $z$. 
Then, one finds from (\ref{BPhieq1}) that $\Xi(x,y)$ obeys 
Liouville's equation $\Delta \Xi +k e^{2\Xi}=0$, where $k$ is a separation constant. 
This implies that the two-dimensional space 
$\D s_2^2=e^{2\Xi(x,y)}(\D x^2+\D y^2)$ 
is maximally symmetric with sectional curvature $k$, which can be taken  $k=0, \pm 1$ 
without loss of generality. It follows that the eqs.~(\ref{BPhieq}) can be solved in full generality and 
the final solution reads 
\begin{align}
\label{RNTNAdS}
\D s^2 &= -f(z) \left(\D t -n \frac{x\D y-y \D x}{1+(k/4)(x^2+y^2)}\right)^2 
-\frac{\D z^2}{f(z)} +\frac{(-z^2+n^2)(\D x^2+\D y^2)}{[1+(k/4)(x^2+y^2)]^2} \,, \nonumber \\
A&= f(z)  \left(\D t -n \frac{x\D y-y \D x}{1+(k/4)(x^2+y^2)}\right)
+\frac{k\ell }4\frac{x\D y -y \D x}{1+(k/4)(x^2+y^2)} \,. 
\end{align}
Here $f(z)=B(z)^2-E(z)^2$ with 
\begin{align}
\label{EB_RNTNAdS}
B=- \left(\frac{n}\ell +\frac{Q z+nP}{-z^2+n^2}\right) \,, \qquad 
E=- \frac z\ell +\frac{Pz+nQ}{-z^2+n^2} \,, 
\end{align}
and the magnetic charge $P$ must obey a Dirac quantization condition,
\begin{align}
\label{}
P=-\frac{k \ell^2+4n^2}{2\ell} \,. 
\end{align}
In particular, the function $\phi$ is given by
\begin{align}
\label{Phi_RNTNAdS}
e^{2\phi}= \frac{f(z)(-z^2+n^2)}{[1+(k/4)(x^2+y^2)]^2} \,. 
\end{align}
The coordinate transformation 
\begin{align}
\label{}
x+iy =\frac {2}{\sqrt k} \tan \left(\frac{\sqrt k}2\theta\right) e^{i\varphi}
\end{align}
brings the metric into a more familiar form, for which the ${\rm U}(1)$ symmetry 
$\partial_\varphi$ is manifest. Note that the coordinate $z$ takes values in $\mathbb R$, 
since there is no restriction on the sign of $f(z)$. 

When the Maxwell field is self-dual, the electric charge 
also obeys the quantization condition
\begin{align}
\label{RNAdS_SD}
Q=P=-\frac{k \ell^2+4n^2}{2\ell } \,. 
\end{align}
In this case, the metric can be written into the 
Przanowski-Tod form \eqref{metr_Prz_Tod} with $z+n=-\ell^2/w$.  
One can easily deduce the explicit expression of $H$ and $u$ 
from (\ref{EB_RNTNAdS}) and (\ref{Phi_RNTNAdS}), 
and check that they satisfy the Toda equation (\ref{Todaeq}).

In \cite{Martelli:2012sz}, the Euclidean supersymmetric Reissner-Nordstr\"om-Taub-NUT 
solution was discussed.  The authors studied the integrability conditions for the Killing spinor to conclude that the self-dual case cannot be supersymmetric unless one additional condition is imposed. 
The loophole is that they worked with the gauge potential obtained by the self-dual limit of the non-self-dual gauge potential.  In fact there is no reason why they should coincide. For example,  the normalization of the Maxwell field is undetermined in the self-dual case since a constant rescaling of the Maxwell field does not affect the field equations. Actually, by appropriately rescaling their gauge potential (with a suitable gauge
transformation preserving \eqref{gauge}), one can write the metric in the Przanowski-Tod-form in
Euclidean signature.

\subsection{Null class}

Let us next discuss the $\Lambda <0$ case where $f=B^2-E^2$ vanishes, i.e., $E=\pm B$. 
This kind of category does not appear in Euclidean signature. As we will demonstrate in appendix~\ref{app:null_class}, the only allowed possibility in this class is that $E=B=0$  and 
$U^\mu$ is a nonvanishing null vector. 
  
Setting $E=B=0$, the algebraic constraints imply\footnote{Due to $W\cdot U=W\cdot W=0$, $W$ is
also parallel to $U$ with a complex proportional factor. Since the differential relation of $W$ fails to
give useful information, we do not consider it here.}
\begin{align}
i_U \Phi = i_U\star \Phi =0\,, \qquad U \we V=0 \,. 
\label{AdS_iUPhi}
\end{align}
Together with the differential constraint
$\D U=-(2/\ell)\star \Phi$, the pseudovector 
$U$ turns out to be hypersurface-orthogonal,
$U \we \D U=0$, hence we can introduce two functions $H$ and $u$ such that 
\begin{align}
U=-H^{-1} \D u \,. 
\end{align} 
If we define the dual coordinate $v$ 
by $U^\mu=(\partial/\partial v)^\mu $, 
$v$ is the Killing coordinate and describes the affine parameter of the null geodesics, 
i.e., the spacetime is a plane-fronted wave. 
Since $V$ is proportional to $U$, the condition $\D V=0$ 
determines the proportional factor as 
\begin{align}
V_\mu =\kappa (u) H U_\mu \,,
\end{align}
where $\kappa=\kappa (u)$ is a function of $u$ only. 
Now introduce a local coordinate system such that
\begin{align}
\D s^2=- H^{-1} \D u (2 \D v-G\D u +2\beta_m \D x^m) + H^{2\alpha} e^{2\phi}(\D x^2-\D w^2) \,, 
\end{align}
where $\alpha$ is a constant introduced for later convenience. $x^m=(x, w)$ 
are the 2-dimensional coordinates orthogonal to $U$. All metric functions 
$(H, G, \beta _m, \phi)$ depend only on $u $ and $x^m$. 

Defining the tetrad
\begin{align}
e^+= H^{-1} \D u \,, \qquad e^- = \D v-\frac 12 G \D u +\beta \,, \qquad 
e^1= H^\alpha e^\phi \D x \,, \qquad 
e^2=H^\alpha e^\phi \D w \,,
\end{align}
with $e^+\we e^-\we e^1\we e^2$ to have positive orientation, 
(\ref{AdS_iUPhi}) constrains the two-form $\Phi$ to be of the form
\begin{align}
\Phi=\Phi_{+i} e^+ \we e^i \,, \qquad 
\star \Phi=\epsilon_{ij} \Phi_{+}{}^j e^+ \we e^i \,, 
\end{align}
where $\epsilon_{12}=-\epsilon_{21}=1$ and its indices $i,j$ are raised and lowered by
$\eta_{ij}={\rm diag}(1,-1)$. 
Inserting this into the relation $\D U=-(2/\ell)\star \Phi$, one gets
\begin{align}
\label{AdS_Null_dUeqsol}
\Phi_{+i}=\frac 12 \ell e^{-\phi} H^{-(1+\alpha)} \epsilon_{ij}\partial^j H  \,,
\end{align}
where $\partial_i=(\partial_x, \partial_w)$. 

Since the Maxwell field satisfies $i_UF=0$, it is restricted to be of the form
\begin{align}
F=F_{+i} e^+ \we e^i +\frac 12 F_{ij}e^i \we e^j \,. 
\end{align}
Note that we do not have a self-dual metric in the null class. 
Due to $i_U\star F=(1/\ell)V$, one has $F_{12}=-(1/\ell)H\kappa$. Plugging this into \eqref{nabla_V}
leads to
\begin{align}
H^{\alpha +2}e^\phi\partial_u (H\kappa )-\ell \epsilon^{ij}F_{+i} \partial_j H =0\,,
\label{AdS_Null_diffeqV}
\end{align}
where $\epsilon^{12}=-\epsilon^{21}=1$. 
The Maxwell equation and the Bianchi identity impose
\begin{align}
\label{AdS_Null_MaxBia}
&0 = \partial_u (H^{1+2\alpha}e^{2\phi}\kappa)-\ell \epsilon^{ij}\partial_j (H^{\alpha -1}e^\phi F_{+i})\,, \\
&0 = \ell \partial^i (H^{\alpha-1}e^\phi F_{+i})-\kappa (\partial_x \beta_y-\partial_y \beta_x)\,.   
\end{align}
The trace of \eqref{AdS_Phieq} gives $\kappa=0$, and the remaining set of equations reads (although these are
not exhaustive)
\begin{subequations}
\begin{align}
&0 = \Box H -\frac 3{2 H}\partial_i H\partial^i H +\frac 4{\ell^2}e^{2\phi} H^{2+2\alpha} \,, 
\label{AdS_Null_DiffeqPhi1}\\
&0 =\partial_i H\partial^i H -\frac 4{\ell^2}e^{2\phi}H^{2+2\alpha} \,,  \label{AdS_Null_DiffeqPhi2}
\\
&0 = \partial_i H(\partial_x \beta_y -\partial_y \beta_x)-2 e^{3\phi}H^{3\alpha+2} \epsilon_{ij}
\partial_u (e^{-\phi} H^{-(1+\alpha)}\partial^j H)\,, \label{AdS_Null_DiffeqPhi3}
\end{align}
\end{subequations}
where $\Box=\partial_x^2-\partial_w^2 $. 
(\ref{AdS_Null_MaxBia}) together with $\kappa=0$ implies that there exists a 
function $\ma F=\ma F(u,x,w)$ such that 
\begin{align}
\label{AdS_Null_Fsol1}
H^{\alpha-1}e^\phi F_{+i}=\partial_i \ma F \,, \qquad 
\Box \ma F=0 \,. 
\end{align}
Substituting into (\ref{AdS_Null_diffeqV}), it turns out that $\ma F$ 
is functionally dependent on $H$ with $u$-dependence, i.e., 
$\ma F=\ma F (u, H)$. 
Compatibility of (\ref{AdS_Null_DiffeqPhi1}) and (\ref{AdS_Null_Fsol1})
yields 
\begin{align}
\ma F= \ell \varphi'(u) H^{1/2} +\varphi_0(u) \,, 
\label{AdS_Null_Fsol}
\end{align} 
where $\varphi$ and $\varphi_0$
are arbitrary functions of $u$. Since $\varphi_0$ does not contribute to the field strength, 
we can set $\varphi_0=0$ without loss of generality. 
We also obtain 
\begin{align}
\label{AdS_Null_Heq}
\Box H^{1/2}=0 \,.
\end{align}
Since $H^{1/2}$ obeys the wave equation on 2-dimensional Minkowski space, 
we can exploit the conformal rescaling 
$u_*\equiv x-w \mapsto \ti u(u_*)$, $v_*\equiv x+w \mapsto \ti v(v_*)$ to achieve 
\begin{align}
\label{}
 H =h(u)
 (x/\ell)^2 \,, 
\end{align}
where 
$h(u)$ is a function of $u$. Putting all together, 
the metric now reads 
\begin{align}
\D s^2 =\frac{\ell^2}{x^2} \left[
-h^{-1}(u) \D u (2\D v-G\D u+2\beta)+\D x^2-\D w^2
\right] \,. 
\end{align} 
We can set $h=1$ by rescaling $u\mapsto U(u)$.  
This amounts to setting $\phi=0$ with $\alpha=-1/2$. 
Eq.~(\ref{AdS_Null_DiffeqPhi3}) implies the existence of a function $W=W(u, x,w)$ such that
\begin{align}
\beta_m=\partial_m W \,. 
\end{align}
We can set  $W=0$ by $v\mapsto v-W$ accompanied by a redefinition of $G$.   
Finally, the ($++$) component of Einstein's equations provides an equation for $G$,
\begin{align}
\label{AdS_Null_Geq}
\Box G -\frac 2{x}\partial_x G+\frac{4x^2}{\ell^2}
 \varphi'(u) ^2 =0 \,. 
\end{align}
To summarize, the solution in the null class reduces to 
\begin{align}
\label{AdS_null_sol}
\D s^2=\frac{\ell^2}{x^2} \left[
- 2 \D v\D u +G\D u^2+\D x^2-\D w^2
\right] \,, \qquad 
A= \varphi(u)\D x \,, 
\end{align}
where $G=G(u,x,w)$ evolves according to (\ref{AdS_Null_Geq}). 

Let us next investigate if the constraints obtained thus far ensure the 
existence of a Killing spinor. 
As a consequence of the projection condition
\begin{align}
\gamma^+\epsilon=0 \,, 
\end{align}
the Killing spinor is $v$-independent in the gauge $i_U A=0$. 
The above projection implies $\gamma^{+-}\epsilon=-\epsilon $, 
which breaks half of supersymmetry. 
The $w$-component of the Killing spinor equation reads 
\begin{align}
\left[\partial_w +\frac 1{2x}\gamma_2 (1-\gamma_1) \right] \epsilon =0 \,, 
\end{align}
which can be solved as 
\begin{align}
\label{}
\partial_w \epsilon=0 \,, \qquad \gamma_1 \epsilon=\epsilon \,. 
\end{align}
The remaining equations are 
\begin{align}
\left(\partial_x -\frac i\ell \varphi 
\right)\epsilon=0 \,, \qquad 
\left(\partial_u -\frac {ix}\ell \varphi'(u) \right) \epsilon =0 \,. 
\end{align}
This is solved by
\begin{align}
\label{AdS_Null_KS}
\epsilon=\exp \left(\frac{ix}\ell \varphi(u) \right) \epsilon_0 \,, 
\end{align}
where $\epsilon_0$ is a constant spinor obeying 
\begin{align}
\gamma^+\epsilon_0=0 \,, \qquad \gamma_1\epsilon_0=\epsilon_0 \,. 
\end{align}
Thus the solution preserves one quarter of supersymmetry.

\subsubsection{Null K\"ahler structure}

Let us consider the two-forms
\begin{align}
\label{NKS_omega}
\omega_1\equiv k_1 F \,, \qquad \omega_2 \equiv k_2 \Phi \,,  
\end{align}
where the functions $k_1, k_2$ are normalization factors
given by $k_1=\ell/(x\varphi'(u))$ and $k_2=\ell^2 x$. 
In terms of these, we define
\begin{align}
\label{AdS_null_J}
\omega_\pm\equiv \omega_1\pm\omega_2\,, \qquad 
J_{\pm\mu}{}^\nu \equiv g^{\nu\rho}\omega_{\pm \rho \mu}=-\left(
\begin{array}{cccc}
 0 & 0 &  1& \pm 1    \\
 0 &  0  & 0 & 0 \\
  0 & 1 & 0 & 0 \\ 
  0& \mp 1 & 0& 0 
\end{array}
\right)\,. 
\end{align}
In the last equality, we work in the coordinate basis ($u , v, x,w$).
It follows that 
$\omega_\pm$ are (anti-)self-dual two-forms, $\star\omega_\pm=\pm \omega_\pm$,
and $J_\pm$ satisfies \cite{Dunajski:2001ea}
\begin{align}
\label{}
J_{\pm\mu}{}^\rho J_{\pm\rho}{}^\nu=0 \,. 
\end{align}
One can also check that the Nijenhuis tensor constructed from $J_\pm$ vanishes, and thus $J_\pm$
is integrable. Let us consider the conformally rescaled metric
\begin{align}
\label{}
\hat g_{\mu\nu}=(x/\ell)^2  g_{\mu\nu}\,,
\end{align}
which is just the metric in the square bracket in (\ref{AdS_null_sol}). Then, one can verify that
$\hat\nabla_\mu J_{\pm\nu}{}^\rho=0$, i.e., the tensor (\ref{AdS_null_J}) defines an integrable null
K\"ahler structure  for $\hat g_{\mu\nu}$. The supersymmetric solution (\ref{AdS_null_sol}) in the null class with $\Lambda<0$ is thus a conformally null K\"ahler manifold. This feature does not arise in Lorentzian
nor Euclidean signatures.


\section{Positive $\Lambda$}
\label{sec:dS}

We shall now discuss the $\Lambda=3L^{-2}>0$ case. Here $L$ corresponds to the inverse `Hubble parameter'.  In Lorentzian signature, this class of theory arises
as `fake supergravity'. Fake supersymmetric solutions have recently attracted some interest, since they
contain black hole geometries embedded in an expanding
universe \cite{Nozawa:2010zg,Chimento:2012mg,Chimento:2014afa}. For a classification of
supersymmetric solutions in fake supergravities
see \cite{Grover:2008jr,Meessen:2009ma,Gutowski:2009vb,Grover:2009ms}.

The Killing spinor equation \eqref{KSE} now reads 
\begin{subequations}
\begin{align}
\hat \nabla_\mu \epsilon &\equiv \left(\nabla_\mu 
+\frac i4 F_{\nu\rho}\gamma^{\nu \rho} \gamma_\mu 
-\frac 1L A_\mu -\frac i{2L}\gamma_\mu \right)\epsilon = 0\,, 
\label{dS_KS}
\\
\overline{\hat \nabla_\mu \epsilon}&=\overline{\nabla_\mu\epsilon}
+\bar \epsilon \left[\frac{i}2 (F_{\mu\nu}+\star F_{\mu\nu}\gamma_5)\gamma^\nu 
-\frac{1}L A_\mu +\frac i{2L} \gamma_\mu\right] \,, \\
\hat \nabla_\mu (\epsilon^TC^{-1}) &=\nabla_\mu (\epsilon^TC^{-1})
+\epsilon^TC^{-1} \left[\frac i2 (F_{\mu\nu}+\star F_{\mu \nu}\gamma_5 )\gamma^\nu 
-\frac 1L A_\mu +\frac i{2L}\gamma_\mu \right]\,.
\end{align}
\end{subequations}
This gives the differential relations
\begin{subequations}
\begin{align}
\nabla_\mu E&=\frac 2L A_\mu E -\star F_{\mu\nu}U^\nu  \,, \label{dS_dE}\\
\nabla_\mu B&=\frac 2L A_\mu B +F_{\mu\nu}U^\nu +\frac 1LU_\mu\,, \label{dS_dB}\\
\nabla_\mu V_\nu &= \frac 2L A_\mu V_\nu -\frac 1L \Phi_{\mu\nu}+F_{(\mu}{}^\rho\Phi_{\nu)\rho}
-\star F_{(\mu}{}^\rho \star \Phi _{\nu)\rho}\,, \label{dS_dV}\\
\nabla_\mu U_\nu &= \frac 2L A_\mu U_\nu -\frac 1L B g_{\mu\nu}-B F_{\mu\nu}-E \star F_{\mu\nu}\,,\label{dS_dU}\\
\nabla_\mu \Phi_{\nu\rho}&= \frac 2L A_\mu \Phi_{\nu\rho} 
+\frac 2Lg_{\mu[\nu}V_{\rho]}+2 F_{\mu [\nu}
V_{\rho]}-V_\mu F_{\nu\rho}-2 g_{\mu[\nu}F_{\rho]\sigma} V^\sigma \,, \label{dS_dPhi}
\end{align}
\end{subequations}
as well as
\begin{subequations}
\begin{align}
\nabla_\mu W_\nu &= \frac 2L A_\mu W_\nu - \frac {1}L \Psi_{\mu\nu}+F_{(\mu}{}^\rho\Psi_{\nu)\rho}
-\star F_{(\mu}{}^\rho \star \Psi _{\nu)\rho}\,,\label{dS_dW}\\ 
\nabla_\mu \Psi_{\nu\rho }&= 
 \frac 2L A_\mu \Psi_{\nu\rho} 
+\frac 2Lg_{\mu[\nu}W_{\rho]}+2 F_{\mu [\nu}
W_{\rho]}-W_\mu F_{\nu\rho}-2 g_{\mu[\nu}F_{\rho]\sigma} W^\sigma\,.
\end{align}
\end{subequations}

The Killing spinor equation (\ref{dS_KS}) is invariant under the $\mathbb R$ 
gauge transformations 
\begin{align}
\label{Rgauge}
A\mapsto A+\D \chi \,, \qquad \epsilon \mapsto \exp(\chi/L) \epsilon \,,
\end{align}
where $\chi $ is a real function. Under \eqref{Rgauge}, all bilinear quantities 
are rescaled by $\exp(2\chi/L)$. 

As in the $\Lambda<0$ case, we analyze the supersymmetric solutions separately 
depending on the behavior of $f=B^2-E^2$. 

\subsection{Non-null class}

Assuming $f\equiv B^2-E^2$ is nonvanishing,  
eqs.~(\ref{dS_dE}) and (\ref{dS_dB}) can be solved to give the Maxwell field
\begin{align}
F=f^{-1} \left[ U \we \left(\D B-\frac {2}{L}BA\right)-\star \left\{
U \we \left(\D E-\frac {2}{L}EA\right)
\right\}\right]\,. \label{dS_F}
\end{align}
As in the $\Lambda<0$ case, the two-form $\Phi$ can be expressed in terms of the other bilinears 
as (\ref{Phi-EBUV}).  Hence the differential constraint of $\Phi$ is automatically satisfied. 

Defining the triple of one-forms $V^i\equiv ({\rm Re}W, {\rm Im}W, V)$, one sees that they follow the same type of differential relations (\ref{dS_dV}) and (\ref{dS_dW}) together with algebraic relations. 
It follows that $V^i$ satisfies 
\begin{align}
\mas L_U V^i= \frac{2}L (i_UA-B) V^i\,,  
\label{dS_Vi_Leiderv}
\end{align}
which suggests to work in the gauge 
\begin{align}
\label{dS_gauge}
U^\mu A_\mu=B \,. 
\end{align} 
In what follows, this gauge condition is assumed. 

Let us introduce a coordinate system such that
\begin{align}
\D s^2 =-f(\D t+\omega)^2 + f^{-1} h_{mn }\D x^m \D x^n \,,
\end{align}
where 
\begin{align}
\label{}
\D s^2 (h) =h_{mn}\D x^m \D x^n= (V^1)^2 +(V^2)^2 -(V^3)^2 \,, 
\end{align}
and $U=\partial/\partial t$. 
Eq.~(\ref{dS_Vi_Leiderv}) implies that base space $\D s^2(h)$ is $t$-independent, while
$f$ and $\omega$ in general can depend on time.
In the gauge (\ref{dS_gauge}),  the gauge potential $A_\mu$ can be decomposed into 
\begin{align}
A =-\frac B f U +\ma B \,, \qquad i_U \ma B=0\,. 
\label{dS_Adec}
\end{align}
Namely, $\ma B_m =A_m -B \omega_m$ as in the $\Lambda<0$ case. (\ref{dS_dE}), (\ref{dS_dB}) and (\ref{dS_dU}) imply then
\begin{align}
\mas L_U \ma B=0\,,
\end{align}
hence $\ma B$ is also time-independent. 
The differential relations for $V^i$ give
\begin{align}
\D V^i =\frac 2L \ma B \we V^i -\frac {2E}{L f} \star (U \we V ^i) \,.   
\end{align}
In the gauge (\ref{dS_gauge}),  we have
\begin{align}
\label{}
\mas L_U (f^{-1} E)=0 \,, \qquad 
\mas L_U (f^{-1} B)=-\frac 1 L \,. 
\label{dS_EBfinv}
\end{align}
Now (\ref{dS_EBfinv}) implies that the function $f^{-1}E $ depends only on the base space coordinates. It
can be set to a constant $\varepsilon =f^{-1}E$ by using the residual gauge freedom 
$A_m \mapsto A_m +\partial_m \chi (x^n )$ of the type (\ref{Rgauge}) preserving the gauge condition (\ref{dS_gauge}). Hence 
\begin{align}
\label{}
\D V^i =\frac 2L\ma B \we V^i -\frac {2\varepsilon}L \star _h V^i \,. 
\label{dS_GT}
\end{align}
Here $\star _h$ is the Hodge dual of the base space. 
Without loss of generality, 
the constant $\varepsilon$ can be scaled to $1$ or $0$. 
In the former case, one finds from (\ref{dS_GT}) that the 
base space defines a three-dimensional (Lorentzian) Gauduchon-Tod structure~\cite{GT}. 
Since the base space function arising from the integration for $B$ 
in  (\ref{dS_EBfinv}) can be set to zero by using the 
freedom $t\mapsto t+g (x^m)$, it follows that 
\begin{align}
E=f \varepsilon \,, \qquad 
B= -\frac tL f \,, \qquad f= \frac{1}{-\varepsilon^2+t^2/L^2}\,.
\end{align}
Using (\ref{dS_F}) and (\ref{dS_Adec}), 
the consistency condition $F=\D A$ gives rise to an equation for $\ma B$,
\begin{align}
\D \ma B -\frac {2\varepsilon}L \star _h \ma B =0 \,. 
\label{dS_eq_calB}
\end{align}
In the $\varepsilon\ne 0$ case, $\ma B$ fulfills the 
divergence-free condition $\D \star_h \ma B=0$. 

Viewing $U=-f(\D t+\omega)$ as a 1-form, (\ref{dS_dU}) provides an equation for the twist form, 
\begin{align}
\D \omega =-\frac 2{Lf} U \we A +\frac{2}{f^2} \star [U \we (E \D B -B\D E)] \,, 
\label{dS_omegaeq}
\end{align}
which implies  $\mas L_U \omega =(2/L) \ma B$, thereby
\begin{align}
\label{dS_omega}
  \omega =\frac 2L \ma B t +\varpi \,, 
\end{align}
where $\varpi$ is a $t$-independent one-form on the base space. 
Inserting (\ref{dS_omega}) back into (\ref{dS_omegaeq}), 
we get 
\begin{align}
\label{dS_domegaeq}
   \D \varpi =\frac 2L \varpi \we \ma B +\frac{2\varepsilon}L 
   \star_h \varpi \,. 
\end{align}
The Maxwell equation leads to 
\begin{align}
\D \star _h \varpi =0 \,. 
\label{dS_Maxeq}
\end{align}
In the $\varepsilon\ne 0$ case, this is already assured by 
the integrability of (\ref{dS_domegaeq}). 

We now show that 
the equations obtained above are also sufficient for the existence of a Killing spinor. 
With the projection condition $i\gamma^0\epsilon=f^{-1/2}(B-E \gamma_5)\epsilon$, 
the time component of the Killing spinor equations becomes
\begin{align}
\hat \nabla_ t \epsilon =\left[\partial_t -\frac 1{2L}(B+E \gamma_5) \right] \epsilon =0\,. 
\end{align}
Due to $\partial_t(B\pm E)=\frac 1{2L}(B\pm E)^2$, 
this equation can be solved as 
\begin{align}
\epsilon =\sqrt{B+E}\zeta^++\sqrt{B-E}\zeta^- \,, 
\end{align}
where $\zeta^\pm=\pm \gamma_5\zeta^\pm $ are time-independent chiral spinors.
Substituting this into the spatial components of the Killing spinor equation, one obtains
\begin{align}
D_m \zeta^\pm \mp \frac{i}L \left(\frac{\varepsilon}2\hat \gamma_m +\epsilon_{mnp}[h]\hat \gamma^n \ma B^p\right)\zeta ^\mp=0 \,. 
\label{dS_KSeq_space}
\end{align}
Since the spin connection for the (Lorentzian) Gauduchon-Tod space
is given by 
\begin{align}
\Omega_{mij}[h] =-\frac 4L \ma B_{[n } h_{p]m}V_i{}^n V_j{}^p
+\frac{\varepsilon}L\epsilon_{ijk}V^k{}_m\,, 
\end{align}
and noting $i\gamma^0\zeta^\pm =\zeta^\mp$, 
the solutions of (\ref{dS_KSeq_space}) are given by constant spinors. 
Hence we arrive at 
\begin{align}
\epsilon =\left(\sqrt{B+E}+i\gamma^0\sqrt{B-E}\right)\frac{1+\gamma_5}2 \epsilon_0 \,, 
\end{align}
where $\epsilon_0$ is a constant Dirac spinor. 
It therefore turns out that the solution preserves at least half of the supersymmetries.

\subsubsection{$\varepsilon=0$ case}

For $\varepsilon=0$ we can simplify the equations and the metric can be obtained explicitly.
Eq.~(\ref{dS_eq_calB}) implies that $\ma B$ can be expressed 
in terms of a local scalar $\psi$ as $\ma B=\D \psi$. 
Using the gauge freedom, we can set $\psi=0$. 
From (\ref{dS_domegaeq}), $\varpi=-L \D H$ for some 
base space function $H$, while (\ref{dS_GT}) leads to $V^i=\D x^i $ with $x^i=(x,y,z)$.
After shifting $t\mapsto t+L H $, we get
\begin{align}
\D s^2=- U^{-2}\D t^2+U^2 (\D x^2+\D y^2-\D z^2) \,, 
\end{align}
where 
\begin{align}
U=\frac tL +H (x,y,z)\,, \qquad (\partial_x^2+\partial_y^2- \partial_z^2) H=0 \,. 
\end{align}
The wave equation of $H$ is a consequence of the Maxwell equation (\ref{dS_Maxeq}). 
This is an analytic continuation of the Kastor-Traschen solution~\cite{Kastor:1992nn}. 

\subsubsection{Self-dual solution}

The self-dual solution $F=\star F$ 
appears only for $\varepsilon\ne 0$. It is easy to verify that this is realized when 
\begin{align}
\varpi=2\varepsilon \ma B \,. 
\end{align}
In this case, the metric can be written as 
\begin{align}
\label{dS_SDsol}
\D s^2=e^{\tau/L}[-H^{-1}(\D \tau+2\ma B)^2+H\D s^2_{\rm GT}]\,, 
\end{align}
where 
\begin{align}
\tau =L\ln \left(\frac{t}L+\varepsilon \right)\,, \qquad H=e^{\tau/L}-2\varepsilon\,,
\end{align}
while $\ma B$ satisfies \eqref{dS_eq_calB}. Notice that the Euclidean version of \eqref{dS_SDsol} was obtained in \cite{Dunajski:2010zp}. A straightforward computation shows that the Weyl tensor for the metric (\ref{dS_SDsol}) (as well as for its $\Lambda<0$ analogue \eqref{metr_Prz_Tod})
is also self-dual. This is actually a consequence of the Killing spinor equation, as is shown
in appendix \ref{sec:sd}.

As an example, let us consider the Wick-rotated Berger sphere as the base space:
\begin{align}
\D s^2_{\rm GT}=\frac{L^2c^2}4[-c^2 (\sigma_3^R)^2+(\sigma_1^R)^2+(\sigma_2^R)^2 ]\,, 
\label{Goedel}
\end{align}
where $c $ is a constant ($0<c\le 1$) and 
$\sigma_i^R$ are the left-invariant  ${\rm SL}(2, \mathbb R)$ one-forms 
\begin{align}
\label{}
    \sigma_1^R&=  -\sin\psi \D \theta+\cos\psi \sinh\theta \D \phi\,, \nonumber   \\
    \sigma_2^R&= \cos\psi \D \theta+\sin\psi \sinh\theta \D \phi \,, \\
    \sigma _3^R&= \D \psi+\cosh\theta \D \phi \,,\nonumber 
\end{align}
satisfying $\D \sigma_i^R=-\frac 12 \epsilon_i{}^{jk}\sigma_j^R \we \sigma_k^R $, where 
indices are raised and lowered by $\eta_{ij}=\eta^{ij}={\rm diag}(1,1,-1)$. The base space 
(\ref{Goedel}) admits the symmetry ${\rm SL}(2,\mathbb R)\times {\rm U}(1)$. 
The Einstein-Weyl triad is given by
\begin{align}
\label{}
   V^1 &=\frac 12 cL[c \sigma_1^L+\sqrt{1-c^2}(\cosh\theta \sigma_2^L-\sinh\theta \sin\phi \sigma_3^L)]\,,\nonumber    \\
 V^2   &= \frac 12 cL[c\sigma_2^L-\sqrt{1-c^2}(\cosh\theta \sigma_1^L+\sinh\theta \cos\phi\sigma_3^L)]\,, \\
 V^3&= \frac 12 cL[c\sigma_3^L -\sqrt{1-c^2} \sinh\theta (\sin\phi \sigma_1^L+\cos\phi \sigma_2^L )] \,, \nonumber   
\end{align}
where 
\begin{align}
\label{}
  \sigma_1^L  &=\sin\phi \D \theta -\cos\phi \sinh\theta \D \psi \,, \nonumber   \\
   \sigma_2^L& =  \cos\phi \D \theta +\sin\phi \sinh\theta \D \psi \,, \\
   \sigma_3^L&=\D \phi+\cosh\theta \D \psi \,.\nonumber 
\end{align}
When $c=1$ the base space reduces to ${\rm AdS}_3$, for which there appears 
an additional ${\rm SL}(2,\mathbb R)$  symmetry to generate 
${\rm SO}(2,2)\simeq {\rm SL}(2,\mathbb R)\times {\rm SL}(2,\mathbb R)$. 
One may thus regard $c$ as the deformation parameter of the bi-invariant group manifold. 
Actually, the metric (\ref{Goedel}) describes the three-dimensional G\"odel universe (see
e.g.~\cite{Gibbons:2011sg}) where $c$ is related to the energy density of the rigidly rotating dust. 

In this case, we have 
\begin{align}
\label{}
\ma B= \frac 12 c\sqrt{1-c^2}L \sigma_3^R \,, \qquad 
\varpi = 2 \ma B \,. 
\end{align}
Following \cite{Dunajski:2013qc}, let us shift the coordinate $\psi$ according to
\begin{align}
\label{}
\D \psi \mapsto \D \psi+ \frac{4\sqrt{1-c^2}L^2 (t+L)}{c[-c^2(t^2-L^2)^2+4L^2(c^2-1)(t+L)^2]}\D t \,.
\end{align}
We thus obtain
\begin{align}
\label{}
\D s^2 = \frac{c^2\D t^2}{4\Delta(t) } +(cL)^2 \Delta(t) (\sigma_3^R)^2 +
\frac 14c^2 (t^2-L^2)
[(\sigma_1^R)^2+(\sigma_2^R)^2 ] \,,
\end{align}
with
\begin{align}
\label{}
\Delta (t)=\frac{c^2(t^2-L^2)^2-4L^2(c^2-1)(t+L)^2}{4L^2(t^2-L^2)} \,. 
\end{align}
By changing to
\begin{align}
\label{}
\hat z=\frac 12 c t \,, \qquad \hat t=cL \psi \,, \qquad n=\frac 12 cL \,, 
\end{align}
one verifies that the metric corresponds to the self-dual limit of the $k=-1$
Reissner-Nordstr\"om-Taub-NUT-de~Sitter metric, which is obtained by taking $\ell \mapsto i L $ in
(\ref{RNTNAdS}) under the constraint (\ref{RNAdS_SD}).

\subsection{Null class}

Let us finally discuss the $\Lambda>0$ case with $E=\pm B$. As for $\Lambda<0$, 
the permitted case occurs only when  $U^\mu $ is a nonvanishing null Killing field with $E=B=0$. 
See appendix~\ref{app:null_class} for details. 

Setting $E=B=0$, it immediately follows from (\ref{dS_dU}) that $U$ is hypersurface-orthogonal, 
hence one can introduce functions $H$ and $u$ such that
\begin{align}
U=-H^{-1} \D u \,. 
\end{align}
We define the dual null coordinate $v$ as 
\begin{align}
\label{}
U^\mu=(\partial/\partial v) ^\mu \,. 
\end{align}
Working in the gauge
\begin{align}
\label{dS_gauge_null}
U^\mu A_\mu=0 \,,
\end{align}
one obtains $U^\nu \nabla_\nu U^\mu=0$, that is, 
the dual coordinate $v$ is an affine parameter of the null geodesics. This means that 
$H$ is independent of $v$. Since $U$ is not Killing, we do not have a plane-fronted wave. 
Rather, the metric is contained in a more general class called the Kundt family, which 
admits a twist-free, non-expanding null geodesic congruence.  

As in the case $\Lambda<0$, we introduce local coordinates as
\begin{align}
\label{}
\D s^2=-H^{-1}\D u(2 \D v-G\D u+2\beta _m \D x^m)+H^{2\alpha}e^{2\phi} (\D x^2-\D w^2) \,, 
\end{align}
and take the tetrad
\begin{align}
e^+= H^{-1} \D u \,, \qquad e^- = \D v-\frac 12 G \D u +\beta \,, \qquad 
e^1= H^\alpha e^\phi \D x \,, \qquad 
e^2=H^\alpha e^\phi \D w\,.
\end{align}
At this stage, the metric components ($G, \beta_m, \phi$) can depend on all coordinates ($u , v, x,w$). 

Writing $V=\kappa U$ for some function $\kappa$, the trace of \eqref{dS_dV} implies that $\kappa$ is
$v$-independent. Plugging this into the antisymmetric part of (\ref{dS_dV}) yields  
\begin{align}
\label{dS_null_Phisol}
\Phi=-\frac L2 \D \kappa \we U \,. 
\end{align}
(\ref{dS_dU}) leads to
\begin{align}
\label{dS_null_diffeqUsol}
\partial_v\phi=0 \,, \qquad 
A_u =-\frac L 4\partial_v G \,, \qquad 
A_i=-\frac L2 \partial_i \ln H \,, \qquad 
\partial_v \beta _i =-\partial_i \ln H \,, 
\end{align}
where $x^i=(x,w)$ with $\partial_i=\partial/\partial x^i$. Since we are using the gauge 
(\ref{dS_gauge_null}), the above conditions determine the field strength as
\begin{align}
\label{dS_F_ex}
F=\frac L4H \partial_v^2 G e^+ \we e^- +
H^{1-\alpha}e^\phi \left[
\frac{L}4(\partial_i \partial_v G-\partial_v^2 G \beta_i)-\frac 12 \partial_u \partial_i (\ln H)
\right] 
e^+ \we e^i \,. 
\end{align}
Note that there exist no solutions with (anti-)self-dual field strength.
(\ref{dS_dE}) is automatically satisfied, while (\ref{dS_dB}) imposes $i_U\star F=0$, 
which implies then $\partial_v^2 G=\frac 4{L^2}H^{-1}$. 
Since $H$ is $v$-independent, this equation can be solved to give 
\begin{align}
\label{}
 G=\frac 2{L^2H} v^2+G_1 (u,x,w) v+G_0(u,x,w) \,. 
\end{align}
Taking into account the $v$-dependence of $\beta $ given in (\ref{dS_null_diffeqUsol}), 
the rescaling of the null coordinate $v\mapsto H v$ together with a 
redefinition of $G_{1,2}$ allows to set $H=1$.
With this choice,  insertion of (\ref{dS_F_ex}) into the Maxwell equations gives
\begin{align}
\label{}
\partial^i (L^2 \partial_i G_1 -4 \beta_i) -8 e^{2\phi}\partial_u \phi=0 \,. 
\end{align}
Eq.~(\ref{dS_dV}) reduces to 
\begin{align}
\label{dS_null_Veqsol}
4e^{2\phi}\partial_u \kappa -(L^2\partial_i G_1-4\beta_i )\partial^i \kappa =0\,,
\end{align}
while (\ref{alg_Phisq}) yields
\begin{align}
\label{dS_null_Phisqeq}
0&=L^2 \partial_i \kappa \partial^i \kappa +4e^{2\phi}(1+\kappa^2 ) \,. 
\end{align}
Finally, \eqref{dS_dPhi} together with (\ref{dS_null_Phisol}) gives rise to 
\begin{subequations}
\begin{align}
0&=\partial_i \partial_j \kappa+\eta_{ij}\partial^k \kappa \partial_k \phi -2\partial_{(i}\kappa
\partial_{j)}\phi +\frac 4{L^2} e^{2\phi}\kappa \eta_{ij} \,, \label{dS_null_Phidiffeq1}\\
0&= e^{2\phi}\partial_u (e^{-2\phi}\partial_i \kappa )+\frac{2\kappa}{L^2}
( L^2 \partial_i G_1 
-4\beta_i)  +e^{-2\phi}\epsilon_{ij}\partial^j \kappa (\partial_x\beta_y-\partial_y \beta_x) \,. 
\label{dS_null_Phidiffeq2}
\end{align}
\end{subequations}
Taking the trace of (\ref{dS_null_Phidiffeq1}) and using (\ref{dS_null_Phisqeq}), 
we have 
\begin{align}
\label{}
\Box ({\rm arctan}\kappa) =0 \,. 
\end{align}
By conformally rescaling $u_*\equiv x-w\mapsto\ti u(u_*)$, $v_*\equiv x+w\mapsto\ti v(v_*)$, we can set 
\begin{align}
\kappa =\tan [\kappa_0 (u) w/L]  \,, 
\end{align}
hence (\ref{dS_null_Phisqeq}) yields
\begin{align}
\label{}
e^{2\phi}=\frac{\kappa_0^2(u)}{4 \cos^2 [\kappa_0 (u) w/L ]}\,.  
\end{align}
Multiplying (\ref{dS_null_Phidiffeq2}) by $\partial^i \kappa$ and using (\ref{dS_null_Veqsol}),
(\ref{dS_null_Phisqeq}), one gets
\begin{align}
\partial_u \left[e^{2\phi}(1+\kappa^2)\right]=0 \,. 
\end{align}
It turns out that $\kappa_0 $ is $u$-independent, i.e., 
$\kappa_0\equiv 1$. From (\ref{dS_null_Phidiffeq2}), 
$\beta_i$ can be obtained as 
\begin{align}
\label{}
\beta_i=\frac L4 \partial_i G_1 \,. 
\end{align}
By shifting $v\mapsto v+\frac {L^2}4G_1$, $\beta_i$ can be made to vanish. It follows that
the line element and the gauge field read
\begin{align}
\label{dS_null_sol}
\D s^2=- \D u \left[2 \D v-\left(2\frac{v^2}{L^2}+G_0(u,x,w) \right)\D u\right]+\frac {\D x^2-\D w^2}{4\cos^2 (w/L)}  \,, 
\qquad 
A= -\frac vL\D u \,. 
\end{align}
The constant $u, v$ slice describes ${\rm dS}_2$. 
 Finally, the ($++$) component of Einstein's equations imposes 
\begin{align}
\label{}
\Box G_0=0 \,. 
\end{align}
When $G_0=0$, the metric (\ref{dS_null_sol}) reduces to ${\rm dS}_2\times {\rm dS}_2$. 
Hence it can be interpreted as a traveling wave on the background ${\rm dS}_2\times {\rm dS}_2$. 
Note also that one cannot take the pure gravitational limit $F_{\mu\nu}=0$, since then the differential
constraint (\ref{dS_dB}) gives rise to an inconsistency. 

Let us move on to solve the Killing spinor equation under the conditions obtained above. 
Using the projection $\gamma^+\epsilon=0$, it is straightforward to check that
\begin{align}
\label{}
\partial_u \epsilon = \partial_v \epsilon =0 \,, 
\end{align}
and
\begin{align}
\label{}
\left(\partial_x -\frac 1{2L}\tan \frac{w}L \gamma_{12} -\frac i{2L}
\sec \frac wL \gamma_1 \right) \epsilon =0 \,, \qquad 
\left(\partial_w -\frac i{2L} \sec \frac wL\gamma_2 \right) \epsilon=0 \,, 
\end{align}
These equations can be solved as 
\begin{align}
\epsilon=
\frac{1}{\sqrt{1-\hat w^2}} (1+ i \hat w \gamma_2) \left(
\cos \frac x{2L}+i \gamma_1 \sin \frac x{2L}
\right) \epsilon_0 \,, 
\end{align}
where $\hat w \equiv \tan \frac w{2L}$ and 
$\epsilon_0$ is a constant spinor with 
$\gamma^+\epsilon_0=0$. 
Hence the solution preserves half of the supersymmetry.

Notice that eq.~\eqref{dS_dU} together with $E=B=0$ implies
\begin{equation}
\nabla_\mu U_\nu = B_\mu U_\nu\,, \qquad B_\mu \equiv \frac2L A_\mu\,, \label{recurrent}
\end{equation}
i.e., the null vector $U$ is recurrent; its direction remains invariant under parallel transport.
In the Lorentzian case this means that the holonomy of $\nabla$ is contained in $\text{Sim}(2)$.
To see what happens for Kleinian signature, we follow the discussion in section 2 of \cite{Gibbons:2007zu}.
First of all, there is a gauge freedom in \eqref{recurrent}, since $U_\nu$ and $\tilde U_\nu=\Omega U_\nu$
describe the same null direction field. Under such a rescaling the recurrence form changes as
\begin{equation}
B \mapsto \tilde B = B + \D\ln\Omega\,.
\end{equation}
Antisymmetrizing the gauge-transformed version of \eqref{recurrent} yields
\begin{equation}
\D\tilde U = \tilde B\wedge\tilde U\,, \label{dtildeU}
\end{equation}
and thus $\tilde U\wedge\D\tilde U=0$. By Frobenius' theorem, there exist therefore two functions
$f,u$ such that $\tilde U=f\D u$. Using the rescaling freedom one may set $f=1$, hence
\begin{equation}
\tilde U = \D u\,, \qquad \D\tilde U = 0\,.
\end{equation}
Plugging this into \eqref{dtildeU} leads to $\tilde B=\eta\tilde U$ for some function $\eta$, and so
the gauge-transformed version of \eqref{recurrent} becomes
\begin{equation}
\nabla_\mu\tilde U_\nu = \eta\tilde U_\mu\tilde U_\nu\,.
\end{equation}
The definition of the Riemann tensor
\begin{equation}
[\nabla_\mu,\nabla_\rho]\tilde U_\nu = R_{\nu\sigma\mu\rho}\tilde U^\sigma
\end{equation}
then implies that
\begin{equation}
R_{\nu\sigma\mu\rho}\tilde U^\sigma = \left[\tilde U_\rho\nabla_\mu\eta - \tilde U_\mu
\nabla_\rho\eta\right]\tilde U_\nu\,. \label{RiemU}
\end{equation}
In our case we have $U=-H^{-1}\D u=-\D u$, so $\Omega=-1$, and $B=(2/L)A=(-2v/L^2)\D u$,
$\tilde B=B$. The function $\eta$ is thus given by $\eta=-2v/L^2$. Moreover, $\tilde U=\tilde U_+ e^+$
with $\tilde U_+=1$, and \eqref{RiemU} simplifies to
\begin{equation}
-R_{\nu -\mu\rho} = \left[\tilde U_\rho\nabla_\mu\eta - \tilde U_\mu\nabla_\rho\eta\right]\tilde U_\nu\,.
\end{equation}
Setting $\nu=i$ gives then
\begin{equation}
R_{i-\mu\rho} = 0\,,
\end{equation}
which leaves the four independent components ${\cal R}_{+-}$, ${\cal R}_{+i}$ and ${\cal R}_{12}$
of the curvature two-form ${\cal R}_{ab}=\frac12 R_{ab\mu\nu}\D x^\mu\wedge\D x^\nu$.
As one easily shows, this means that the holonomy is contained in $\text{Sim}(1)\times\text{Sim}(1)$
$\subset$ $\text{SL}(2,\mathbb R)\times\text{SL}(2,\mathbb R)\simeq {\rm SO}(2,2)$. 
Note that $\text{Sim}(1)$ is the two-dimensional 
subgroup of the Lorentz group $\text{SO}(2,1)$ in $2+1$ dimensions generated by $H,D$ satisfying
$[D,H]=H$.

\section{Concluding remarks}
\label{sec:summary}

In this paper we have explored some geometric properties of spaces admitting Killing spinors in
minimal $N=2$ gauged supergravity with Kleinian signature. We classified the geometries according to the
sign of the cosmological constant and the causal nature of a vector field constructed from the Killing spinor. Spaces with two time directions are important in the context of twistor space. Some exact supersymmetric
self-dual solutions obtained in this paper might be interesting testgrounds for this purpose. Also, it would
be interesting to explore the F-theory interpretation of the solutions constructed here.  

Using the bilinear approach, we revealed some new features which are also present in the Lorentzian and Euclidean cousins, but unnoticed in the literature.  We first pointed out the utility of using supplementary
bilinears (\ref{WPsi}), which considerably simplify the analysis of classification by the bilinear technique.
An intriguing result is the appearance of the generalized monopole equation~(\ref{gen_monopole}), reminiscent of Einstein-Weyl spaces. For the self-dual subclass, this is indeed the case, since the base
space of the form $\D s_3^2=\pm \D z^2+e^u(\D x^2+\D y^2)$ together with the continuous Toda
equation defines an Einstein-Weyl structure~\cite{tod_p}. However, the interpretation in the non-self dual
case is not yet clear, and might be related to some generalization of Einstein-Weyl structures that have
not been discussed in the math literature so far.

We also clarified new aspects of supersymmetric geometries intrinsic to Kleinian signature. The bilinear
vector field can be timelike, spacelike or null, in contrast to the Lorentzian and Euclidean cases. This
implies that a broad class of supersymmetric solutions is allowed compared to the previous studies. It was
shown that the null class of the $\Lambda<0$ case gives rise to an integrable null K\"ahler structure. In
order to define this, the Maxwell field plays an essential role [see (\ref{NKS_omega})]. Hence the null
K\"ahler structure does not occur in the purely gravitational case. It would be interesting to see if null
K\"ahler structures also arise for supersymmetric solutions in higher dimensions.

\section*{Acknowledgements} 

We would like to thank Sigbj{\o}rn Hervik for useful comments. 
This work is partially supported by JSPS and INFN.


\appendix
\section{Spin connection}
\label{app:spincon}

In this appendix we summarize some useful formulae for the spin connection used in the body of the text. 

\subsection{Non-null class}

Consider the metric with ($2,2$) signature of the form
\begin{align}
\D s^2=-f (\D t +\omega)^2 +f^{-1} h_{mn}\D x^m \D x^n \,, 
\end{align}
where the base space metric $h_{mn}$ is assumed to be $t$-independent, 
while $f$ and $\omega$ depend on $t$ and $x^m$. This class of metric encompasses the 
non-null class with both signs of $\Lambda$. 
Choosing the tetrad
\begin{align}
e^0=f^{1/2}(\D t+\omega)\,, \qquad e^i =f^{-1/2} \hat e^i \,,
\end{align}
with 
\begin{align}
\label{}
 h_{mn}=\eta_{ij} \hat e^i{}_m \hat e^j{}_n \,, \qquad \eta_{ij}={\rm diag}(1,1,-1)\,,
\end{align}
the spin connection $\Omega_{abc}=\Omega_{a[bc]}$ reads
\begin{align}
\Omega_{00i} &= f^{1/2}\left(-\frac 12 \ma D_m f+f\dot \omega_m\right)\hat e_i{}^m\,, \\
\Omega_{0ij}&=f^{5/2} \ma D_{[m}\omega_{n]} \hat e_i{}^m \hat e_j{}^n \,, \\
\Omega_{k0i}&= f^{1/2}\left(f\ma D_{[m}\omega_{n]}+\frac 12 f^{-2}\dot f h_{mn} \right)\hat e_k{}^m \hat e_i{}^n \,, \\
\Omega_{kij}&=f^{1/2} \left(\Omega_{mij}[h]-h_{m[n}\ma D_{p]}\ln f \hat e_i{}^n \hat e_j{}^p\right) \hat e_k{}^m \,,
\end{align} 
where $i,j,k$ are frame components, while $m,n,..$ are coordinate components of the base space. 
The dot denotes a differentiation with respect to $t$, $\Omega_{mnp}[h]$ 
is the spin connection of the base space, and $\ma D_m=\partial_m -\omega_m \partial_t $. 
Using the spin connection given above and assuming the projection condition
$i\gamma^0 \epsilon=f^{-1/2}(B-E \gamma_5)\epsilon$,  
the supercovariant derivative of the Einstein-Maxwell theory,
\begin{align}
\label{SCD_EM}
\hat \nabla_\mu \epsilon =\left(
\partial_\mu +\frac 14 \Omega_{\mu ab}\gamma^{ab}+\frac{i}4 F_{\nu\rho} \gamma^{\nu\rho} \gamma_\mu \right)\epsilon \,, 
\end{align}
can be decomposed into
\begin{align}
\hat \nabla_t \epsilon =& \left[
\partial_t -\frac{i}{2}f^{1/2}\gamma^i \left\{
F_{0i}+\frac{1}{2f}\left({B\partial_m f+E\Omega_m}-{B(\dot f \omega_m
 +2 f\dot \omega_m )}\right)\hat e_i{}^m
\right\}
\right. \nonumber \\
& \left. +\frac{1}{2}f^{1/2}\gamma_5\gamma^i  \left\{\star F_{0i}+
\frac{1}{2f} \left(-E\partial_m f -B\Omega_m +E(\dot f \omega_m+2 f \dot
 \omega_m )\right)\hat e_i{}^m 
\right\}  
\right]\epsilon \,,\\
\hat \nabla_i\epsilon  =&f^{1/2}\hat e_i{}^m \left[
D_m -\omega_m \partial_t 
+\frac{1}{2f}\hat e^j{}_m(F_{0j}-\star F_{0j}\gamma_5)(B-E\gamma_5)
+\frac{i}{4f^{5/2}}\dot f \hat \gamma_m (B-E\gamma_5)
\right.\nonumber \\ & \left. 
+\frac{i}{2f^{1/2}}\epsilon_{mnp}[h]\hat \gamma^n  h^{pq}
\left\{\star F_{0j} \hat e^j{}_q-\frac 1{2f}(B \Omega_q+E\ma D_q f) 
\right\}
\right.\nonumber \\ & \left. 
+\frac{i}{2f^{1/2}}\epsilon_{mnp}[h]\hat\gamma^n  \gamma_5 h^{pq} \left\{
F_{0j}\hat e^j{}_q+\frac 1{2f}(E \Omega_q+B\ma D_qf)
\right\}
\right]\epsilon \,,
\end{align}
where $\hat \gamma^m=\gamma^i\hat e_i{}^m$, while $D_m$ and $\epsilon_{mnp}[h]$ are the
covariant derivative and the volume element of the base space.  Moreover we have defined
\begin{align}
\Omega_m =\epsilon_{mnp}[h]h^{nq}h^{pr}\ma D_{[q}\omega_{r]}\,, 
\end{align}
which describes the twist of $U^\mu=(\partial/\partial t)^\mu$, i.e., 
 $\Omega_\mu=\epsilon_{\mu\nu\rho\sigma}U^\nu\nabla^\rho U^\sigma$.

\subsection{Null class} 

The metric in the null class discussed in the body of the text can be written universally as 
\begin{align}
\D s^2 = H^{-1}[-\D u (2 \D v-G\D u) +e^{2 \phi} (\D x^2-\D w^2)]\,. 
\end{align}
Here $G=G(u,v,x^m)$, whereas $H$ and $\phi$ depend only on $x^m=(x,w)$.  
We choose the null tetrad
\begin{align}
\label{}
e^+= H^{-1} \D u \,, \qquad e^-=\D v-\frac 12 G \D u\,, \qquad 
e^i= H^{-1/2}e^{\phi} \D x^i \,, \qquad 
\eta_{ab}=[-\sigma_1, \sigma_3] \,. 
\end{align}
The nonvanishing components of the spin connection $\Omega_{abc}$ are given by
\begin{align}
\Omega_{++-}&=\frac 12 H\partial_ v G \,, \qquad 
\Omega_{++i}=\frac 12 e^{-\phi}H^{3/2} \partial_i G \,, \nonumber \\
\Omega_{+-i}&=\Omega_{-+i}=\Omega_{i+-}=\frac 12 H^{-1/2} e^{-\phi }\partial_i H \,,   \\
\Omega_{ijk}&=e^{-\phi}H^{-1/2} \eta_{i[j} (-\partial_{k] } H+2 H\partial_{k]} \phi ) \,, \nonumber 
\end{align}
where $\partial_i=(\partial_x, \partial_w )$. 
Suppose that the only nonvanishing components of the Maxwell field
are $F_{+-}$ and $F_{+i}$. Then, using the projection condition $\gamma^+\epsilon=0$, 
the supercovariant derivative (\ref{SCD_EM}) decomposes into 
\begin{subequations}
\begin{align}
\label{}
\hat \nabla_+ \epsilon &= \left[H\left(\partial_u+\frac 12 G \partial_v\right)
-\frac 14 H \partial_v G +\frac 14 e^{-\phi} H^{-1/2}\partial_i H \gamma^{-i}
+\frac i2 F_{+-}\gamma^- +i F_{+i} \gamma^i 
\right]\epsilon \,, \\
\hat \nabla_- \epsilon &=\partial_v \epsilon \,, \\ 
e_i{}^\mu \hat \nabla _\mu \epsilon &= H^{1/2}e^{-\phi} \left[
\partial_i -\frac 14 \partial_i (\ln H)+\frac 1{4H} (-\partial_j H+2 H\partial_j \phi)
\gamma_i{}^j -\frac i2 H^{-1/2} e^\phi F_{+-} \gamma_i 
\right] \epsilon \,.
\end{align}
\end{subequations}
These expressions are useful in order to compute explicitly the Killing spinor.

\section{Holonomy of the base manifold for $\Lambda<0$}
\label{app_holonomy}

Supergravity solutions admitting Killing spinors are typically fibrations over base manifolds with
reduced holonomy, at least in the timelike case. For instance, in minimal ungauged $N=2$, $D=4$
supergravity, the base is flat \cite{Tod:1983pm} and thus has trivial holonomy. This is still true if
one couples the theory to vector multiplets \cite{Meessen:2006tu}. In five-dimensional minimal
ungauged supergravity, the base manifold is hyper-K\"ahler \cite{Gauntlett:2002nw}, while in the gauged
case it is K\"ahler \cite{Gauntlett:2003fk}. One might therefore ask whether (\ref{metric_base}) has
reduced holonomy as well. It turns out that this is actually the case, but for a torsionful (or alternatively
nonmetric) connection. To see this, start from the first Maurer-Cartan structure equation for a
three-dimensional spacetime with tangent space metric $\eta_{ij}=\mathrm{diag}(1,1,-1)$,
\begin{align}
\D\hat e^i + {\Gamma^i}_j\wedge\hat e^j = T^i\,, \label{1stMC}
\end{align}
$T^i$ being the torsion two-form. Suppose that the connection $\Gamma$ has holonomy
$\mathrm{U}(1)\subset\mathrm{SO}(2,1)$. This means that $\Gamma$ must have the form
\begin{align} \label{conn_red}
\Gamma_{ij} = \left(\begin{array}{ccc} 0 & \alpha & 0 \\ -\alpha & 0 & 0 \\ 0 & 0 & 0\end{array}\right)\,,
\end{align}
where $\alpha$ is a one-form. Under the additional assumption $T^3=0$, (\ref{1stMC}) implies
$\D \hat e^3=0$, and thus $\hat e^3=\D z$ for some function $z$. The remaining two eqs.~of (\ref{1stMC})
can be written as
\begin{align}
\D\hat e^{\pm} \mp i\alpha\wedge\hat e^{\pm} = T^{\pm}\,, \label{de_pm}
\end{align}
with the complex forms $\hat e^{\pm}\equiv\hat e^1\pm i\hat e^2$, $T^{\pm}\equiv T^1\pm iT^2$.
Let us suppose further (the reason for this will become clear in a moment) that the torsion satisfies
also $\hat e^+\wedge T^+=0=\hat e^-\wedge T^-$. Then, (\ref{de_pm}) implies
$\hat e^{\pm}\wedge\D\hat e^{\pm}=0$, and thus there exist complex functions $\eta$ and $w$ such
that $\hat e^+=\eta\D w$, $\hat e^-=\bar\eta\D\bar w$. Plugging this back into (\ref{de_pm})
leads to
\begin{align} \label{eq_eta}
\eta,_{z} = i\eta\alpha_z + T^+_{zw}\,, \qquad \eta,_{\bar w} = i\eta\alpha_{\bar w} + T^+_{\bar w w}\,.
\end{align}
If we define $\eta=\eta_0e^{i\rho}$, $T^+_{zw}=\eta(a+ib)$, with $\eta_0,\rho,a,b$ real, and use the
$\mathrm{U}(1)$ gauge freedom
\begin{align}
\hat e^i \mapsto {M^i}_j\hat e^j\,, \qquad {M^i}_j = \left(\begin{array}{ccc} \cos\theta & \sin\theta & 0 \\
-\sin\theta & \cos\theta & 0 \\ 0 & 0 & 1 \end{array} \right)\,,\qquad \alpha \mapsto \alpha - \D\theta\,,
\end{align}
which preserves the form (\ref{conn_red}), to set $\rho=0$, the first eq.~of (\ref{eq_eta}) gives
$\alpha_z=-b$ and $\partial_z\ln\eta_0=a$, and hence $\eta_0=\exp\int a\D z$. This leads to the
metric
\begin{align}
\D s^2 = \eta_{ij}\hat e^i\hat e^j = -\D z^2 + e^{2\int a\D z}\D w\D\bar w\,,
\end{align}
which is exactly what we have (cf.~(\ref{metric_base}) and set $w=x+iy$). Moreover, from (\ref{eq_phip})
we see that the torsion component $a$ is given by $a=-(F_++F_-)/2$.
We can thus interpret the base space (\ref{metric_base}) as a manifold of reduced holonomy
$\mathrm{U}(1)\subset\mathrm{SO}(2,1)$ with nonzero torsion. Note that the holonomy with respect to
the Levi-Civita connection is not reduced. 
A similar case occurs in five-dimensional minimal de~Sitter supergravity, where the (timelike)
supersymmetric solutions are fibrations over a hyper-K\"ahler manifold with torsion
(HKT) \cite{Grover:2008jr}.

Reduced holonomy is equivalent to the existence of parallel tensors, the simplest example being
the reduction of $\mathrm{GL}(D,\mathbb{R})$ to $\mathrm{SO}(D)$ if the metric is covariantly
constant, $\nabla g=0$. In our case, the corresponding parallel tensor is just the vector $\partial_z$,
which is easily seen to be covariantly constant w.r.t.~the torsionful connection ${\Gamma^i}_j$.

Actually, we can see (\ref{1stMC}) directly in (\ref{nabla_V}) and (\ref{nabla_W}), which imply (after
projection onto the base)
\begin{align} \label{dVdW}
\D V = 0\,, \qquad \D W = \frac{2i}{\ell}\left(A - B\omega - \frac{iE}{f}V\right)\wedge W\,.
\end{align}
From (\ref{local_scal}) it is clear that $V=\hat e^3$, $W=\hat e^+$, and thus the second eq.~of (\ref{dVdW})
is exactly (\ref{de_pm}), if we identify
\begin{align}
T^+ = \frac{2i}{\ell}\left(A - B\omega - \frac{iE}{f}V - \frac{\ell}{2}\alpha\right)\wedge\hat e^+\,.
\end{align}
Note that the one-form $\alpha$ is undetermined at this stage, since we are free to absorb $\alpha$
either into the connection or into the torsion.

It is interesting to see what happens if we trade the torsion for nonmetricity, which can of course always
be done. We wish to rewrite (\ref{dVdW}) in the form
\begin{align}
\D\hat e^i + \hat\Gamma^i_{\,\,j}\wedge\hat e^j - \frac12\nu\wedge\hat e^i = 0\,, \label{1stMC_nonmetr}
\end{align}
where $\hat\Gamma$ is a metric connection ($\hat\Gamma_{ij}=-\hat\Gamma_{ji}$), and $\nu$
denotes a one-form. The Weyl connection $\hat\Gamma^i_{\,\,j}-\frac12\nu\delta^i_{\,\,j}$ is nonmetric,
but has zero torsion. (\ref{1stMC_nonmetr}) is invariant under Weyl rescalings
\begin{align}
\hat e^i\mapsto e^{\psi}\hat e^i\,, \qquad \nu\mapsto\nu + 2\D\psi\,.
\end{align}
One finds that (\ref{dVdW}) can be written in the form (\ref{1stMC_nonmetr}) if the components
$\hat\Gamma^{ij}$ are given by
\begin{subequations}
\begin{align}
\hat\Gamma^{12} &= \frac12(\nu^1\hat e^2 - \nu^2\hat e^1) + \frac{2E}{\ell f}\hat e^2\,, \\
\hat\Gamma^{13} &= \frac12(\nu^1\hat e^3 - \nu^3\hat e^1) + \frac{2E}{\ell f}\hat e^3\,, \\
\hat\Gamma^{23} &= \frac12(\nu^2\hat e^3 - \nu^3\hat e^2) + \frac{2}{\ell}(A-B\omega)\,.
\end{align}
\end{subequations}
The gauge field $\nu$ appearing here is of course arbitrary, and there is a priori no reason why one
should choose the one-form $\nu$ of the generalized monopole equation (\ref{gen_monopole}).
It may be that some constraints on the curvature of $\hat\Gamma$ (like e.g.~the Einstein condition)
single out the one-form $\nu$ of (\ref{gen_monopole}), but we did not check this explicitly.

\section{Integrability conditions and equations of motion}
\label{sec:Int}

In this appendix we address the question to what extent the Killing spinor equations imply
the second order equations of motion in neutral signature. First of all, a spinorial equation of the form
\begin{align}
\hat\nabla_\mu\epsilon = \left[\nabla_\mu + c_1\slashed{F}\gamma_\mu + c_2\gamma_\mu
+ c_3 A_\mu\right]\epsilon = 0\,,
\end{align}
where $\slashed{F}\equiv F^{ab}\gamma_{ab}$ and $c_1$, $c_2$, $c_3$ are complex constants, has the
first integrability conditions
\begin{eqnarray}
\left[\hat\nabla_\mu,\hat\nabla_\nu\right]\epsilon &=& \left[\frac14{R^{ab}}_{\mu\nu}\gamma_{ab} + c_1
\left((\nabla_\mu F^{ab})\gamma_{ab}\gamma_\nu - (\nabla_\nu F^{ab})\gamma_{ab}\gamma_\mu
+ \slashed{F}{T^a}_{\mu\nu}\gamma_a\right)\right. \nonumber \\
&& +\: c_2 {T^a}_{\mu\nu}\gamma_a + c_3 F_{\mu\nu} +
4c_1^2\left(4 F^{\rho\lambda}F_{\rho\left[\mu\right.}\!\gamma_{|\lambda|\left.\!\nu\right]}
- F^2 \gamma_{\mu\nu}\right) \nonumber \\
&&\left.+\: 8c_1c_2\left(\star F_{\mu\nu}\gamma_5 - {F_{\left[\mu\right.\!}}^\rho
\gamma_{|\rho|\left.\!\nu\right]}\right) + 2c_2^2\gamma_{\mu\nu}\right]\epsilon = 0\,,
\label{int_cond_KSE}
\end{eqnarray}
${T^a}_{\mu\nu}$ being the components of the torsion two-form. Contracting (\ref{int_cond_KSE})
with $\gamma^\nu$, assuming vanishing torsion\footnote{It would be interesting to relax this.},
and using the Bianchi identities for the Riemann and Faraday tensor as well as the Maxwell equations,
one obtains\footnote{To get (\ref{Einstein_gamma}) one has to choose the coefficients such that
$c_3=-8c_1c_2$ in order to cancel terms linear in the field strength $F$. For the Killing spinor equation
(\ref{KSE}) this is of course satisfied, since $c_1=i/4$, $c_2=\sqrt{-\Lambda/3}/2$,
$c_3=-i\sqrt{-\Lambda/3}$.}
\begin{align} \label{Einstein_gamma}
E_{ab}\gamma^b\epsilon = 0\,, \qquad \mathrm{where}\quad E_{ab} \equiv
R_{ab} + 12c_2^2 g_{ab} + 32c_1^2\left(F_{ac}{F_b}^c - \frac14 F^2 g_{ab}\right)\,.
\end{align}
Multiplying this from the left with $\bar\epsilon$ yields
\begin{align}
E_{ab}V^b=0\,, \label{E_V}
\end{align}
while multiplication with $\bar\epsilon\gamma_5$ gives
\begin{align}
E_{ab}U^b=0\,. \label{E_U}
\end{align}
Finally, hitting (\ref{Einstein_gamma}) from the left with $E_{ac}\gamma^c$ we deduce that
\begin{align}
E_{ac}{E_a}^c = 0\,, \qquad \mathrm{no}\,\,\mathrm{sum}\,\,\mathrm{over}\,\,a\,. \label{E^2}
\end{align}
If there exists an orthonormal frame in which $U$ has only $0$-component and $V$ has only
$3$-component, the eqs.~(\ref{E_V}) and (\ref{E_U}) imply $E_{a0}=E_{a3}=0$. Choosing
$a=A$ (where $A=1,2$) in (\ref{E^2}) one gets then
\begin{align}
\sum_{B=1}^2(E_{AB})^2 = 0 \quad\Rightarrow\quad E_{AB} = 0\,,
\end{align}
and thus all the Einstein equations $E_{ab}=0$ are satisfied\footnote{In the Lorentzian case, generically
it happens that also a part of the Maxwell equations is implied by the Killing spinor equations. This can
be shown by using Killing spinor identities \cite{Bellorin:2005hy}. Perhaps one can shew something
analogous in neutral signature, but we shall not attempt to do this here.}.

In the null case things are a little bit more subtle for Kleinian as compared to Lorentzian signature.
If there exists a null frame $(e^+,e^-,e^1,e^2)$ such that
\begin{equation}
\eta_{ab} = \left(\begin{array}{rrrr} 0 & -1 & 0 & 0 \\ -1 & 0 & 0 & 0 \\ 0 & 0 & 1 & 0 \\ 0 & 0 & 0 & -1
\end{array}\right)\,,
\end{equation}
in which $U=U_+e^+$ (note that this is satisfied for
both signs of $\Lambda$), eq.~\eqref{E_U} gives $E_{a-}=0$. Using this in \eqref{E^2} yields
\begin{equation}
E_{a1}{E_a}^1 + E_{a2}{E_a}^2 = 0\,.
\end{equation}
If we were in Lorentzian signature, this would imply $E_{a1}=E_{a2}=0$, but here one
can only conclude that
\begin{equation}
(E_{a1})^2 - (E_{a2})^2 = 0\,. \label{E^2_null}
\end{equation}
However, we have also the other bilinears at our disposal. For instance, hit \eqref{Einstein_gamma}
from the left with $i\bar\epsilon\gamma^c$ to get
\begin{equation}
E_{ab}\Phi^{cb} + i E {E_a}^c = 0\,. \label{E_Phi}
\end{equation}
In the null cases one has $E=0$ and $\Phi=\Phi_{+2}e^+\wedge e^2$ (with $\Phi_{+2}\neq 0$), and
thus \eqref{E_Phi} boils down to
\begin{equation}
E_{a2}\Phi^{c2} = 0\,, \label{E_Phi1}
\end{equation}
where we used also $E_{a-}=0$. Taking $c=-$, \eqref{E_Phi1} yields $E_{a2}=0$. Plugging this into
\eqref{E^2_null} one obtains that also $E_{a1}=0$. Therefore the only equation of motion one has
to impose is $E_{++}=0$.

Notice finally that
in order to preserve maximal supersymmetry, each coefficient in (\ref{int_cond_KSE}) in terms of the
Clifford basis $\{1, \gamma_5, \gamma_\mu, \gamma_\mu\gamma_5, \gamma_{\mu\nu}\}$
must vanish separately. One finds thus that the maximally supersymmetric geometries with
$\Lambda\ne 0$ are exhausted by constant curvature spacetimes with $F_{\mu\nu}=0$.

\section{Self-duality of the Weyl tensor}
\label{sec:sd}

Here we show that the integrability conditions \eqref{int_cond_KSE}, together with the equations of
motion and the self-duality condition for the electromagnetic field strength $F_{\mu\nu}$, imply that
the Weyl tensor must be self-dual as well. First of all, decompose the Riemann tensor in
\eqref{int_cond_KSE} according to
\begin{equation}
R_{\mu\nu\rho\sigma} = C_{\mu\nu\rho\sigma} + g_{\mu\left[\rho\right.}R_{\left.\sigma\right]\nu}
- g_{\nu\left[\rho\right.}R_{\left.\sigma\right]\mu} - \frac R3 g_{\mu\left[\rho\right.}
g_{\left.\sigma\right]\nu}\,, \label{decomp-Riem}
\end{equation}
and use the equations of motion
\begin{equation}
R_{\mu\nu} = -12c_2^2 g_{\mu\nu} - 32c_1^2\left(F_{\mu\rho}{F_\nu}^\rho - \frac14 F^2 g_{\mu\nu}
\right) = -12c_2^2g_{\mu\nu}\,,
\end{equation}
(where the second step holds due to self-duality of $F$) to eliminate the Ricci tensor and scalar curvature
from \eqref{decomp-Riem}. Then the integrability conditions \eqref{int_cond_KSE} become
\begin{eqnarray}
&&\left[\frac14{C^{ab}}_{\mu\nu}\gamma_{ab} + c_1
\left((\nabla_\mu F^{ab})\gamma_{ab}\gamma_\nu - (\nabla_\nu F^{ab})\gamma_{ab}\gamma_\mu
\right) + c_3 F_{\mu\nu}(1 - \gamma_5)\right. \nonumber \\
&&\quad\left.+\: c_3{F_{\left[\mu\right.\!}}^\rho\gamma_{|\rho|\left.\!\nu\right]}
+ 4c_1^2\left(4 F^{\rho\lambda}F_{\rho\left[\mu\right.}\!\gamma_{|\lambda|\left.\!\nu\right]}
- F^2 \gamma_{\mu\nu}\right)\right]\epsilon = 0\,. \label{int_cond_Weyl}
\end{eqnarray}
If we write $\epsilon=\epsilon_++\epsilon_-$, where the chiral spinors $\epsilon_{\pm}$ were
defined in section \ref{selfdual-negLambda}, eq.~\eqref{int_cond_Weyl} splits into a positive chirality
and a negative chirality component. The former reads
\begin{eqnarray}
&&\left[\frac14{C^{ab}}_{\mu\nu}\gamma_{ab} + 4c_1^2\left(4 F^{\rho\lambda}F_{\rho\left[\mu\right.}
\!\gamma_{|\lambda|\left.\!\nu\right]} - F^2 \gamma_{\mu\nu}\right)
+ c_3{F_{\left[\mu\right.\!}}^\rho\gamma_{|\rho|\left.\!\nu\right]}\right]\epsilon_+ \nonumber \\
&&\quad +\: c_1\left((\nabla_\mu F^{ab})\gamma_{ab}\gamma_\nu - (\nabla_\nu F^{ab})\gamma_{ab}
\gamma_\mu\right)\epsilon_- = 0\,. \label{int_cond_Weyl_chir}
\end{eqnarray}
Now, using the self-duality of $F$, the second relation of \eqref{gamma_eps} and
$\gamma_5\epsilon_{\pm}=\pm\epsilon_{\pm}$, one obtains
\begin{equation}
{F_\mu}^\rho\gamma_{\rho\nu}\epsilon_+ = {F_\nu}^\rho\gamma_{\rho\mu}\epsilon_+ +
\frac12 g_{\mu\nu}\slashed{F}\epsilon_+\,,
\end{equation}
and thus ${F_{\left[\mu\right.\!}}^\rho\gamma_{|\rho|\left.\!\nu\right]}\epsilon_+=0$. Moreover,
the same ingredients imply
\begin{displaymath}
(\nabla_\mu F^{ab})\gamma_{ab}\gamma_\nu\epsilon_- = 0\,, \qquad
4 F^{\rho\lambda}F_{\rho\left[\mu\right.}\!\gamma_{|\lambda|\left.\!\nu\right]} -
F^2 \gamma_{\mu\nu} = 0\,, \qquad {C^{ab}}_{\mu\nu}\gamma_{ab}\epsilon_+ =
{C^{-ab}}_{\mu\nu}\gamma_{ab}\epsilon_+\,,
\end{displaymath}
so that \eqref{int_cond_Weyl_chir} boils down to
\begin{equation}
{C^{-ab}}_{\mu\nu}\gamma_{ab}\epsilon_+ = 0\,.
\end{equation}
Contracting this from the left with $i\bar\epsilon_+$ yields
\begin{equation}
{C^{-ab}}_{\mu\nu}\Phi^-_{ab} = 0\,, \label{C-Phi}
\end{equation}
whereas hitting with $i\epsilon^T_+ C^{-1}$ gives
\begin{equation}
{C^{-ab}}_{\mu\nu}\Psi^-_{ab} = 0\,. \label{C-Psi}
\end{equation}
But $\Phi^-$, $\text{Re}\Psi^-$ and $\text{Im}\Psi^-$ form a basis in the space of anti-self-dual
two-forms, as can be seen from the expressions
\begin{equation}
\Phi^- = \frac1{B-E}\left(V\wedge U\right)^-\,, \qquad \Psi^- = \frac{iE}f\left(V\wedge W\right)^-
- \frac Bf\left(U\wedge W\right)^-\,, \label{Phi-Psi-asd}
\end{equation}
that follow from \eqref{Phi-EBUV} and \eqref{Psirel} respectively. Using the orthonormal basis
\begin{equation}
e^0=-f^{-1/2}U\,, \qquad e^1 + i e^2 = f^{-1/2}W\,, \qquad e^3 = f^{-1/2}V\,, 
\end{equation}
\eqref{Phi-Psi-asd} can also be written as
\begin{displaymath}
\Phi^- = (B+E)\left(e^0\wedge e^3\right)^-\,, \quad \text{Re}\Psi^- = (B+E)\left(e^0\wedge
e^1\right)^-\,, \quad \text{Im}\Psi^- = (B+E)\left(e^0\wedge e^2\right)^-\,,
\end{displaymath}
from which it is evident that $\Psi^-$, $\text{Re}\Psi^-$ and $\text{Im}\Psi^-$ are linearly
independent. \eqref{C-Phi} and \eqref{C-Psi} imply therefore
\begin{equation}
{C^{-ab}}_{\mu\nu} = 0\,,
\end{equation}
hence the Weyl tensor is self-dual.
Note that an analogous result was obtained in \cite{Dunajski:2010zp} for Euclidean signature, using
the two-component spinor language. In that case, the positive-definiteness of the metric has been used to conclude the statement.

\section{On the classification of the null class}
\label{app:null_class}

In the body of the text, we classified the supersymmetric solutions in the null class 
under the condition that $U^\mu $ is a null vector with $E=B=0$. We show in this appendix that other possibilities are excluded. 

Suppose $f=B^2-E^2=0$. We take the sign $B=E$ for convenience.  
We have then two possibilities depending on
whether (i) $U^\mu$ identically vanishes or (ii) $U^\mu $ is a null vector. 
Note that in the Lorentzian case, the existence of a Killing spinor immediately implies that the null Killing vector is nonvanishing, hence the case (i) does not arise.  Moreover, 
unlike in the Lorentzian case, two null vectors orthogonal to each other are not necessarily parallel
in neutral signature.  Hence we need to be more careful for the classification of the null class. 
To proceed, we will have to use the differential relations for the bilinears. Thus, we shall discuss the
two cases $\Lambda \gtrless 0$ separately below.

\subsection{Negative $\Lambda$}

Consider first the case (i), where $U^\mu=0$ with $B=E$. 
Then, the algebraic constraints (\ref{alg_starPhiUV}) give $EV_\mu=0$, leading to (i-a) $E=0$
or (i-b) $V_\mu =0$. For (i-a), (\ref{AdS_diff_E}) and (\ref{nabla_U}) imply $V_\mu=\Phi_{\mu\nu}=0$, 
incompatible with a nonvanishing Killing spinor. For (i-b), (\ref{alg_EBPhi}) implies 
$\Phi^+\equiv\frac 12(\Phi+\star\Phi)$ vanishes [the $E=0$ case reduces to (i-a)]. 
From (\ref{nabla_U}) and its Hodge dual, $\Phi_{\mu\nu}=0$. Thus, (\ref{alg_PhistarPhi}) leads to 
an inconsistency again. It follows that there are no supersymmetric solutions in case (i).

Consider next the case (ii), where the vector $U^\mu$ is a nonvanishing null vector. 
We can then write 
\begin{align}
\label{U_dec}
U^\mu =\ti \kappa V^\mu +K^\mu \,, 
\end{align}
where $\ti \kappa$ is some (possibly vanishing) proportionality factor and $K^\mu$ denotes
another null vector which is linearly independent of $V^\mu $ and satisfying 
$K^\mu K_\mu=V^\mu K_\mu=0$. 
From the algebraic relation  (\ref{alg_EBPhi}), 
we have
\begin{align}
\label{Phip_UV_ind}
2E \Phi^+_{\mu\nu}=\epsilon_{\mu\nu\rho\sigma}V^\rho U^\sigma\,.
\end{align}
Here we can consider two possibilities: (ii-a) $E=0$ and (ii-b) $E \ne 0$. 
For (ii-a), (\ref{Phip_UV_ind}) implies that $U^\mu$ and $V^\mu$ are linearly dependent, 
hence $K^\mu=0$. For (ii-b), 
the differential relations (\ref{AdS_diff_E}), (\ref{AdS_diff_B}), (\ref{nabla_U}) together with 
the algebraic relations (\ref{alg_PhiUV}) 
imply then 
\begin{align}
\label{}
V_\mu=\frac {\ell}{4E} \epsilon_{\mu\nu\rho\sigma}U^\nu \nabla^\rho U^\sigma \,,
\label{V_twist_AdS}
\end{align}
i.e., $V_\mu$ is proportional to the twist of the Killing vector $U^\mu$. 
Substituting (\ref{V_twist_AdS}) into (\ref{Phip_UV_ind}), we have 
$\Phi^+=0$, hence $K^\mu=0$. Inserting $\Phi^+=0$ into 
(\ref{alg_PhistarPhi})--(\ref{alg_starPhisq}) we have a contradiction. 
Therefore the case (ii-b) cannot occur and 
the only allowed case of $f=B^2-E^2=0$ for $\Lambda<0$
is that $U^\mu $ is a nonvanishing null Killing field with $E=B=0$.

\subsection{Positive $\Lambda$}

In case (i), we have $EV_\mu=0$. If $V^\mu$ identically vanishes, the differential relation \eqref{dS_dV}
together with \eqref{alg_Phisq}, \eqref{alg_starPhisq} gives $E=B=\Phi=0$, incompatible with the existence of a Killing spinor. If $E=B=0$ with a nonvanishing null vector $V^\mu$, we have 
$\Phi_\mu{}^\rho\Phi_{\nu\rho}=-V_\mu V_\nu$ and $i_V\Phi=i_V\star \Phi=0$.
Since $\Phi$ is antisymmetric, there exists a matrix $S\in\text{SO}(2,2)$ such that
$\Phi_{ab}={S_a}^c Q_{cd}{{S^T}^d}_b$, where
\begin{equation}
Q = \left(\begin{array}{cccc} 0 & \lambda_1 & 0 & 0 \\ -\lambda_1 & 0 & 0 & 0 \\
       0 & 0 & 0 & \lambda_2 \\ 0 & 0 & -\lambda_2 & 0 \end{array}\right)\,. \label{Q}
\end{equation}
Defining $\tilde V^a\equiv V^b {S_b}^a$, the eq.~$i_V\Phi=0$ becomes
$Q_{ab}\tilde V^b=0$, and thus $Q$ must have a zero eigenvalue. Without loss of generality
we assume $\lambda_1=0$. Then, if $\lambda_2\neq 0$, $\tilde V$ has to be of the form
$\tilde V=(\tilde V^0,\tilde V^1,0,0)^T$. 
Using the fact that the volume element 
$\epsilon$ is an invariant tensor under $\text{SO}(2,2)$, i.e.,
$\epsilon^{abcd}{S_a}^e{S_b}^f{S_c}^g{S_d}^h=\epsilon^{efgh}$, 
$i_V\star\Phi=0$ is equivalent to $\star Q_{ab}\tilde V^b=0$. Since $\star Q$ is the same as
\eqref{Q} but with $\lambda_1$ and $-\lambda_2$ interchanged, the latter eq.~implies
$\tilde V^0=\tilde V^1=0$ and thus $V=0$, which contradicts our assumption that the null
vector $V$ is nonvanishing. The other possibility is that also $\lambda_2=0$, but then $\Phi=0$
and from $\Phi_\mu{}^\rho\Phi_{\nu\rho}=-V_\mu V_\nu$ we get again $V=0$. 
Hence the case (i) cannot occur. 

In case (ii), we can decompose the vector $U^\mu $ as (\ref{U_dec}), and 
consider (ii-a) $E=0$ or (ii-b) $E \ne 0$.  
In case (ii-a), (\ref{Phip_UV_ind}) means $K^\mu=0$. 
In case (ii-b), the differential constraints (\ref{dS_dE}), (\ref{dS_dB}),  (\ref{dS_dU}) imply
\begin{align}
\label{dS_U_UVind}
U_\mu =\frac L{2E}\epsilon_{\mu\nu\rho\sigma}U^\nu\nabla^\rho U^\sigma\,.
\end{align}
Substituting this into (\ref{Phip_UV_ind}), we have $\Phi^+=0$. Again, 
this contradicts (\ref{alg_PhistarPhi})--(\ref{alg_starPhisq}), and thus the case (ii-b) cannot arise. 

In conclusion, the admissible case of the null class in either sign of the cosmological constant 
is $E=B=0$ with $U^\mu$ being a nonvanishing null vector.


\end{document}